\documentclass[a4paper,11pt]{article}
\pdfoutput=1

\usepackage{jheppub}

\usepackage[usenames,dvipsnames]{xcolor}
\usepackage[english]{babel}
\usepackage{enumerate,enumitem,float,graphicx,comment}

\usepackage[T1]{fontenc}
\usepackage[utf8]{inputenc}
\usepackage{lmodern}
\usepackage[normalem]{ulem}

\makeatletter
\g@addto@macro\bfseries{\boldmath}
\makeatother

\usepackage{amsmath,amssymb,mathtools}

\usepackage{tabulary}
\newcolumntype{K}[1]{>{\centering\arraybackslash}p{#1}}

\newcommand{\pd}{\partial}

\newcommand{\eps}{\varepsilon}

\newcommand{\LL}{\mathcal{L}}

\newcommand{\mA}{\mathcal{A}}

\newcommand{\OO}{\mathcal{O}}

\newcommand{\spa}{\ \ ,\ \ \ \ }
\newcommand{\CO}{{\cal{O}}}
\newcommand{\CL}{{\cal{L}}}

\newcommand{\ab}{{\underline{a}}}
\newcommand{\bb}{{\underline{b}}}
\newcommand{\cb}{{\underline{c}}}
\newcommand{\db}{{\underline{d}}}

\title{%
  Torsional string Newton--Cartan geometry for non-relativistic strings
}

\author[a]{Leo Bidussi,}
\author[b]{Troels Harmark,}
\author[a]{Jelle Hartong,}
\author[b,c]{Niels A. Obers,}
\author[b]{Gerben Oling}

\affiliation[a]{%
  School of Mathematics and Maxwell Institute for Mathematical Sciences,
  University of Edinburgh,
  \\
  Peter Guthrie Tait Road,
  Edinburgh EH9 3FD, UK
}
\affiliation[b]{%
  The Niels Bohr Institute,
  University of Copenhagen,
  \\
  Blegdamsvej 17,
  DK-2100 Copenhagen Ø,
  Denmark
}
\affiliation[c]{%
  Nordita,
  KTH Royal Institute of Technology and Stockholm University,
  \\
  Hannes Alfv\'ens v\"ag 12, SE-106 91
  SE-106 91 Stockholm,
  Sweden
}

\emailAdd{l.bidussi@sms.ed.ac.uk}
\emailAdd{harmark@nbi.ku.dk}
\emailAdd{j.hartong@ed.ac.uk}
\emailAdd{obers@nbi.ku.dk}
\emailAdd{gerben.oling@nbi.ku.dk}

\abstract{%
  We revisit the formulation of non-relativistic (NR) string theory and its target space geometry.
  We obtain a new formulation in which the geometry contains a two-form field that couples to the tension current and that transforms under string Galilei boosts.
  This parallels the Newton--Cartan one-form that couples to the mass current of a non-relativistic point particle.
  We show how this formulation of the NR string arises both from an infinite speed of light limit and a null reduction of the relativistic closed bosonic string.
  In both cases, the two-form originates from a combination of metric quantities and the Kalb--Ramond field.
  The target space geometry of the NR string is seen to arise from the gauging of a new algebra that is obtained by an İnönü--Wigner contraction of the Poincaré algebra extended by the symmetries of the Kalb--Ramond field.
  In this new formulation, there are no superfluous target space fields that can be removed by fixing a Stückelberg symmetry.
  Classically, there are no foliation/torsion constraints imposed on the target space geometry.
}

\begin{document}
\maketitle
\flushbottom

\section{Introduction}%
\label{sec:intro}
In recent years, non-relativistic (NR) string theory has received considerable attention. Originally, it was formulated as an infinite speed of light limit of a relativistic string in flat space  with a near-critical electric Kalb--Ramond field \cite{Gomis:2000bd,Danielsson:2000gi}. Recently, it has been extended to include a general target space geometry
 \cite{Harmark:2017rpg,Kluson:2018egd,Bergshoeff:2018yvt,Harmark:2018cdl,Gallegos:2019icg,Harmark:2019upf}. Along with this, there are several studies of the target space geometry  \cite{Andringa:2012uz,Bergshoeff:2018vfn,Bergshoeff:2021bmc,Yan:2021lbe}, the beta-functions \cite{Gomis:2019zyu,Gallegos:2019icg,Yan:2019xsf,Bergshoeff:2019pij,Yan:2019xsf}, and its Hamiltonian formalism \cite{Kluson:2018egd,Kluson:2018grx,Kluson:2018vfd,Kluson:2019qgj,Kluson:2019xuo,Kluson:2019ajy}.
The NR string has been related to limits of AdS/CFT \cite{Gomis:2005pg,Harmark:2017rpg,Harmark:2018cdl,Harmark:2019upf,Harmark:2020vll,Fontanella:2021hcb} and double field theory \cite{Berman:2019izh,Park:2020ixf,Gallegos:2020egk,Blair:2020gng,Morand:2021xeq}. Supersymmetric non-relativistic strings have been studied in \cite{Blair:2019qwi}.
The open string sector has been explored in \cite{Gomis:2020fui,Gomis:2020izd}.

In this paper we revisit the target space geometry in the NS-NS sector of the closed NR string
and find a novel formulation, which we call torsional string Newton--Cartan (TSNC) geometry.
To motivate this, we start by reviewing the current state of the art and putting our results in the context of these  previous developments.

String Newton--Cartan (SNC) geometry originates from the gauging of the string Newton--Cartan algebra \cite{Bergshoeff:2018vfn}. One can write down string probe actions that have all the gauge symmetries of SNC geometry, but this requires a foliation constraint to be imposed on the target space. Furthermore, a Kalb--Ramond 2-form can be included in the formulation, but this leads to an overparametrization of the field content as manifested in the appearance of a St\"uckelberg gauge symmetry that allows one to remove some of the target space fields. The new formulation presented in this work resolves these issues as there is no foliation constraint necessary and there is no overparametrization of the field content. For reasons that will be explained below we will refer to this new target space geometry as torsional string Newton--Cartan (TSNC) geometry.

Our formulation also brings the NR string in close analogy with the NR point particle case, and its coupling to torsional Newton--Cartan (TNC) geometry. In particular, we find that the natural generalization of the one-form mass connection $m_\mu$ of the point particle is a two-form gauge field $m_{\mu\nu}$ that in part arises from the Kalb--Ramond two-form.  Connected to this, we find a new underlying non-relativistic  string algebra for the target space geometry in the NS-NS sector, which we refer to as the F-string Galilei (FSG) algebra.

The analysis of this paper emanates from revisiting the construction of NR strings in curved target spacetimes, which has in recent years been actively studied from various points of view.  Up to this point, the Nambu-Goto action plus Wess-Zumino term%
\footnote{See \cite{Harmark:2017rpg,Harmark:2018cdl} for earlier work on non-relativistic strings with zero $B$-field, based on null reduction.} for the NS-NS sector of the NR string that has been considered is \cite{Bergshoeff:2018yvt,Harmark:2019upf}
\begin{equation}
\label{SNC_action}
S=- \frac{T}{2} \int d^2\sigma \Big[ \sqrt{-\tau} \, \tau^{\alpha\beta} \bar{h}_{\mu\nu} + \epsilon^{\alpha\beta} B_{\mu\nu} \Big] \partial_\alpha X^\mu \partial_\beta X^\nu\,,
\end{equation}
where we omitted the dilaton contribution to the action since it will not play a role below.
Here the target space geometry consists of the transverse metric $h_{\mu\nu}$, the longitudinal vielbeine $\tau^A_\mu$ with $A=0,1$ and the two one-forms $m^A_\mu$ such that one has
\begin{equation}
  \bar{h}_{\mu\nu}
  = h_{\mu\nu}
  + \eta_{AB} \left( \tau^A_{\mu} m^B_{\nu} + \tau^A_{\nu} m^B_{\mu} \right),
\end{equation}
with $\mu,\nu=0,1,\ldots,D-1$.
Furthermore, we define the pullbacks $\tau^A_\alpha = \partial_\alpha X^\mu \tau^A_\mu$ and
the resulting worldsheet metric $\tau_{\alpha\beta}=\eta_{AB}\tau^A_\alpha\tau^B_\beta$ is assumed to be Lorentzian,
with its inverse denoted by $\tau^{\alpha\beta}$ and determinant $\tau=\det\tau_{\alpha\beta}$. For simplicity we consider here the Nambu--Goto form rather than the Polyakov action since our focus is on  understanding the target space geometry.

A number of issues arise with the action \eqref{SNC_action}.
First of all, the string Newton--Cartan geometry that has been conjectured as the structure of the target space geometry has a gauge symmetry know as the $Z_A$ gauge symmetry, with parameter $\sigma^A$, which acts as
\begin{equation}
\label{ZAsym}
    \delta m_\mu^A=D_\mu\sigma^A=\partial_\mu\sigma^A-\omega_\mu\varepsilon^A{}_B\sigma^B\,,
\end{equation}
where $\omega_\mu$ is an $SO(1,1)$ connection associated with the Lorentz transformations acting on~$\tau^A_\mu$.
This $Z_A$ gauge symmetry was originally introduced in analogy with the $U(1)$~symmetry appearing in the  Bargmann algebra for the particle case, corresponding to mass conservation \cite{Andringa:2012uz}.
From this analogy, it is clear one should have a commutator of the type
\begin{equation}
\label{GPZcom}
[G_{Aa},P_b] =\delta_{ab} Z_A\,,
\end{equation}
that connects the string boost generators $G_{Aa}$, the transverse translation generators $P_b$ and $Z_A$.
This is the string generalization of the Bargmann commutator $[G_a,P_b]=\delta_{ab}N$ of the Galilei boosts $G_a$ and the translations that gives the $U(1)$ generator $N$.
Indeed, the SNC algebra of \cite{Andringa:2012uz} realizes the commutator~\eqref{GPZcom}, but at the price that in order to close the algebra
(i.e. obey the Jacobi identities) one needs to introduce an additional generator $Z_{AB}=\epsilon_{AB} Z $ in the commutator of two string boosts.
This generator has no corresponding field in the target space geometry, nor an analogue in the particle case.

Secondly, realizing the $Z_A$ symmetry as a symmetry of the action \eqref{SNC_action} imposes a non-trivial constraint $D_{[\mu} \tau^A_{\nu]}=0$ on the target space geometry, where $D_\mu$ is the $SO(1,1)$ covariant connection used in the definition \eqref{ZAsym} of the $Z_A$ symmetry. This can be interpreted as a no-torsion constraint. It has become increasingly clear, from various considerations such as backgrounds with R-R fluxes, supersymmetry and constraints from beta function computations \cite{Harmark:2018cdl,Gomis:2019zyu,Gallegos:2019icg,Yan:2019xsf,Bergshoeff:2019pij,Bergshoeff:2021bmc,Yan:2021lbe}, that this is condition on the target space geometry is too restrictive.

Thirdly, the action is invariant under the Stückelberg transformations \cite{Bergshoeff:2018yvt,Harmark:2019upf}
\begin{equation}
  \bar{h}_{\mu\nu} \rightarrow \bar{h}_{\mu\nu} + 2 C_{(\mu}^A \tau_{\nu )}^B \eta_{AB} \spa B_{\mu\nu} \rightarrow B_{\mu\nu} - 2 C_{[ \mu}^A \tau_{\nu ]}^B \epsilon_{AB},
\end{equation}
for any two one-forms $C^A_\mu$, $A=0,1$.
This freedom amounts to $m_\mu^A\rightarrow m_\mu^A+C_\mu^A$ with an appropriate shift in the $B$-field. This shows that the $m_\mu^A$ field is redundant as one can for instance choose a gauge in which $m^A_\mu=0$. One would expect that, when taking the $c\rightarrow\infty$ limit, it should be possible to arrive at the minimal set of fields needed for the formulation of the theory. Indeed, taking the $c\rightarrow \infty$ limit in the analogous particle case one does not find such auxiliary fields.
On the other hand, in an expansion of the relativistic string in $1/c^2$,
the $m^A_\mu$ fields appear at next-to-leading order in the expansion of the relativistic vielbeine.
They end up being paired with the next-to-leading order part of the Kalb--Ramond two-form in a manner that admits a shift symmetry.
Again, the same happens for a point particle (see \cite{Hansen:2020pqs}). Generalizing the methods developed in  \cite{Hansen:2019pkl,Hansen:2020pqs}  for the $1/c^2$ expansion of general relativity and the coupling of point particles
to non-relativistic geometry,
the systematic $1/c^2$ expansion of the relativistic string is presented in
the recent works \cite{typeIITSNC1,typeIITSNC2}.
In this approach, one does not demand that the leading order action
 cancels, in contradistinction to what has been done so far to construct non-relativistic string actions. We will return to this below, when we juxtapose the  limiting and expansion procedures for both the point particle and string case.

Finally, one can instead obtain the action for a NR string from a null reduction of the relativistic string.
In the particle case, the $c\rightarrow \infty$ limit is known to be equivalent to a null reduction. As we review in Section~\ref{sec:null_reduction}, one finds the following action from the null reduction of the NS-NS sector of the relativistic fundamental string \cite{Harmark:2019upf}
\begin{equation}
\label{null_SNC_action}
S=- \frac{T}{2} \int d^2\sigma \Big[ \sqrt{-\tau} \tau^{\alpha\beta} h_{\mu\nu} + \epsilon^{\alpha\beta} m_{\mu\nu} \Big] \partial_\alpha X^\mu \partial_\beta X^\nu\,,
\end{equation}
where $h_{\mu\nu}$ is the transverse metric, $\tau^0_\mu$ comes from the clock one-form of the null reduction, $\tau^1_\mu$ comes from the Kalb--Ramond field along the null direction and $m_{\mu\nu}$ is a sum of the part of the Kalb--Ramond field transverse to the null direction and a linearly independent part that combines $\tau^1_\mu$ and a one-form $m_\mu$ from the null-reduced metric.
Together, $\tau_\mu^A$, $h_{\mu\nu}$ and $m_{\mu\nu}$ contain the same information as the fields appearing in the theory \eqref{SNC_action} obtained from the $c\to\infty$ limit.
In fact, the action \eqref{SNC_action} includes two extra one-forms $m^A_\mu$ that are not part of the action \eqref{null_SNC_action}.
These extra one-forms precisely account for the auxiliary fields that give rise to the Stückelberg symmetry, and once they have been removed by fixing the Stückelberg the two approaches agree in their field content.

The question we ask and answer in this paper is: can one regard \eqref{null_SNC_action} as the fundamental action of the NR string in the NS-NS sector instead of \eqref{SNC_action}?
If so, this would require a new understanding of the target space geometry, since it is clear that the transverse metric~$h_{\mu\nu}$ is not invariant under string Galilean boosts.
This implies that  one needs to regard~$m_{\mu\nu}$ as part of the geometry, as it will have to transform under these local boosts.

Moreover, with \eqref{null_SNC_action} as the fundamental action, one is forced to regard the $Z_A$ symmetry as a symmetry of the $m_{\mu\nu}$ field itself. But this is actually an advantage, since $m_{\mu\nu}$ precisely has the gauge symmetry $m \rightarrow m+ d \lambda $ for any one-form $\lambda$.
We will show that one can reinterpret the $Z_A$ symmetry as the longitudinal part of this gauge symmetry. This in turn suggests that one can realize this symmetry also when the torsion is non-zero, since the gauge symmetry of $m_{\mu\nu}$ as such does not depend on $\tau^A_\mu$ and $h_{\mu\nu}$.

We will thus show in this paper that the target space geometry of the NS-NS sector of the NR string consists of $\tau^A_\mu$, $h_{\mu\nu}$ and $m_{\mu\nu}$, and we find the underlying algebra that realizes this geometry.
This algebra is analogous to the Bargmann algebra of the particle case as it includes the commutator \eqref{GPZcom} but not the $Z_{AB}$ generators of the SNC algebra.
We show that one can get the action \eqref{null_SNC_action} directly from a $c\rightarrow \infty$ limit.
The new algebra we find is a proper İnönü--Wigner contraction of the algebra underlying the target space of the relativistic string.
We emphasize that the TSNC two-form $m_{\mu \nu}$ appearing in the action \eqref{null_SNC_action} couples to the tension current of the NR string, and transforms non-trivially under string Galilei boosts.
Moreover, for a string that is point-like in all directions except for a compact longitudinal direction $v$ with periodicity $2 \pi R$  along which it winds, the NR string action naturally reduces to that of a NR particle  with mass $2 \pi R T$, where the Newton—Cartan gauge field is the one-form obtained from integrating $m_{\mu v}$ over the spatial direction along the string

As mentioned above, a different though related problem is the $1/c^2$ {\it expansion} of the relativistic string
as opposed to the $c \rightarrow \infty $ {\it limit}.
This is reminiscent of the $1/c^2$ expansion of the point particle case studied in \cite{Hansen:2020pqs}. In \cite{typeIITSNC1,typeIITSNC2} this approach is studied for closed bosonic strings.
In Table~\ref{table:geometries} we summarise the main differences between on the one hand the limit/null reduction approach, which we will refer to as type I, and on the other hand the~$1/c^2$~expansion approach, which we will refer to as type II.%
\footnote{We will only make this distinction here and not in the rest of paper which exclusively deals with the type I case.}
Furthermore, we also highlight the similarities between the particle and string cases for both type I and type II objects.
To aid the reader, we have included a brief review of the type I NR particle in Appendix~\ref{sec:particle}, while the type II NR particle is treated in \cite{Hansen:2020pqs}.

\begin{table}[t]
      \centering
      {\renewcommand{\arraystretch}{1.2}\scalebox{.82}{
      \begin{tabular}{|K{2cm}|K{3cm}|K{5cm}|K{4cm}|K{3cm}|}
      \cline{1-5}
        & origin & target space geometry of probe action & torsion/foliation constraints & important special cases\\
       \cline{1-5}
       NR particle (type I) & $c\rightarrow\infty$ limit of extremal particle/null reduction of massless particle& (type I) TNC geometry:\hspace{2cm} $\tau_\mu$, $h_{\mu\nu}$, $m_\mu$ & none & NC geometry: $d\tau=0$\\
       \cline{1-5}
       NR string (type I) & $c\rightarrow\infty$ limit of string with critical electric field/null reduction of relativistic string & (type I) TSNC geometry:\hspace{2cm} $\tau_\mu^A$, $h_{\mu\nu}$, $m_{\mu\nu}$ & none & SNC geometry: $d\tau^A=\omega\varepsilon^A{}_B\wedge\tau^B$\\
       \cline{1-5}
       NR particle (type II) plus relativistic corrections & $1/c^2$ expansion of uncharged massive relativistic particle& (type II) TNC geometry: \hspace{1cm} LO: $\tau_\mu$ \hspace{4cm} NLO: $\tau_\mu$, $h_{\mu\nu}$, $m_\mu$ \hspace{1.5cm} NNLO: $\tau_\mu$, $h_{\mu\nu}$, $m_\mu$, $\Phi_{\mu\nu}$, $B_\mu$ & dynamically determined by both the EOM of the target space fields and the embedding scalars & NC geometry: $d\tau=0$\\
       \cline{1-5}
       NR string (type II) plus relativistic corrections & $1/c^2$ expansion of uncharged relativistic string & (type II) TSNC geometry: LO: $\tau^A_\mu$ \hspace{3cm} NLO: $\tau^A_\mu$, $h_{\mu\nu}$, $m^A_\mu$ & dynamically determined by both the EOM of the target space fields and the embedding scalars & $d\tau^A=\alpha^A{}_B\wedge\tau^B$ with $\alpha^A{}_A=0$\\
       \cline{1-5}
        \end{tabular}}}
\caption{Overview of the different approaches to NR particles and strings.}\label{table:geometries}
\end{table}

In the latter case, there are finitely many fields at each order in $1/c^2$ and the NR particle is the theory obtained at NLO. The subleading orders describe relativistic corrections. The type I particle couples to a geometry that can be obtained by gauging the Bargmann algebra \cite{Andringa:2010it,Hartong:2015zia}, without imposing any condition on $\tau_\mu$.
Type II TNC geometry can be obtained by gauging the $1/c^2$-expanded Poincaré algebra whose field content up to NNLO was used in \cite{Hansen:2019pkl,Hansen:2020pqs} and whose notation we also use here.
It should be borne in mind that the fields appearing at a certain order in the probe action are in general a subset of the fields appearing at the same order in $1/c^2$-expansion of the bulk gravity action. Another comment is that we here only consider the case of an expansion in even powers of $1/c$.
When one allows for odd powers, more options in how the expansion is done become available \cite{VandenBleeken:2017rij,Ergen:2020yop,Hansen:2020pqs}.
The expansion of the equation of motion of the embedding scalars tells us that $\dot x^\mu (d\tau)_{\mu\nu}=0$ where $\tau_\mu\dot x^\mu>0$. In combination with the LO expansion of the Einstein equation which forces the Frobenius condition $\tau\wedge d\tau=0$ for a codimension-one foliation, this tells us that $d\tau=0$. Hence, the $1/c^2$ expansion of gravity coupled to an uncharged point particle tells us that the geometry must admit an absolute time foliation. This is in stark contrast with the $c\rightarrow\infty$ limit of an extremal particle in a near-critical electric field, whose action is defined for any $\tau_\mu$.

In the present work and in \cite{typeIITSNC1,typeIITSNC2}  the analogous situations for strings is worked out. For the convenience of the reader we have summarised our findings in Table~\ref{table:geometries}. The situation regarding the $1/c^2$-expansion of the relativistic string in an arbitrary metric (but ignoring the Kalb--Ramond field) is as follows.

The $1/c^2$ expansion of the string involves a rescaling by $c$ of two vielbeine, one of which is the timelike vielbein. In the particle case we only rescale the timelike vielbein. If we expand the Einstein equations in a string $1/c^2$-expansion (in the absence of the Kalb--Ramond field and the dilaton) we learn that at LO the Einstein equations force the geometry to admit a foliation of codimension-two leaves. The LO Einstein equations are equivalent to the Frobenius condition $d\tau^A=\alpha^A{}_B\wedge\tau^B$ where $A,B=0,1$ and where $\tau^A$ are the normal 1-forms to the leaves of the foliation. The LO term in the equation of motion of the embedding scalars then constrains the geometry slightly further \cite{typeIITSNC1}. A special case of the target space geometry allowed by the $1/c^2$-expansion is the SNC geometry that one obtains by gauging the SNC algebra (see Appendix~\ref{app:fsnc-gauging} and \cite{Bergshoeff:2018vfn}). To be clear, by SNC geometry we mean a geometry that obeys $d\tau^A=\omega\varepsilon^A{}_B\wedge\tau^B$ and for which we do not include the Kalb--Ramond 2-form.
Finally, there is a string analogue of the particle $c\rightarrow\infty$ perspective which is what the present paper addresses in detail.

This paper is structured as follows. In Section \ref{sec:string_limit} we consider the infinite speed of light limit of the relativistic string in the NS-NS sector. This is in analogy with the particle case reviewed in Appendix~\ref{sec:particle}.
We begin in Section \ref{sec:gaugingKR} by showing how one can formulate the target space geometry of the relativistic string using a gauging of the string Poincaré algebra, which consists of the Poincaré generators plus additional generators that give rise to the Kalb--Ramond field.
In Section \ref{sec:TSNCgeometry} we take the infinite speed of light limit of the relativistic Nambu--Goto action and obtain the NR string with the new torsional SNC target space.
In Section \ref{sec:symmetries_TSNC} we exhibit the symmetries of the TSNC geometry, showing how it can be obtained from the gauging of an underlying string Bargmann algebra, realized as an İnönü--Wigner contraction of the string Poincaré algebra.

In Section \ref{sec:null_reduction} we review how one obtains the NR string Nambu--Goto action \eqref{null_SNC_action} from a null reduction of the relativistic string in the NS-NS sector. Subsequently, in Sections \ref{sec:nullred_boost} and \ref{sec:nullred_ZA}, we exhibit how the string Galilean boost transformations and the one-form gauge symmetry arise from the null reduction point of view.

In Section \ref{sec:conclusion} we present our conclusions and consider the future directions that our novel NR string geometries opens up.
In Appendix \ref{sec:particle} we review the $c\rightarrow\infty$ and null reduction approaches for the point particle. Finally, in Appendix~\ref{app:algebras} we further describe the novel FSG algebra and its gauging as well as a more detailed comparison
to the SNC algebra and its associated geometry.

\subsubsection*{Note added}

While we were in the final stages of writing up this paper we became aware of the paper \cite{Yan:2021lbe} which likewise aims at relaxing the foliation constraint of the NR string's target space, motivated by the quantum theory.

\section{Torsional string Newton--Cartan target space from \texorpdfstring{$c\rightarrow\infty$}{c->infty} limit}
\label{sec:string_limit}

In this section we show how torsional string Newton--Cartan (TSNC) geometry emerges naturally as a target space geometry of a neutral non-relativistic (NR) string resulting from a $c\rightarrow \infty$ limit of the fundamental relativistic string in the NS-NS sector.%
\footnote{In a slight abuse of terminology, NS-NS sector refers here to the metric and Kalb--Ramond field, leaving the dilaton aside, and we briefly comment on its coupling in the conclusion.}

As argued in the Introduction, the longitudinal NS-NS $B$-field plays a crucial role in obtaining the action for non-relativistic strings.
In fact, this is already evident in the original Gomis--Ooguri analysis~\cite{Gomis:2000bd}, which considers non-relativistic strings in a flat target spacetime by taking a near-critical field limit.
Therefore, as a first step in setting up our $c\rightarrow \infty$ limit, we discuss in Section \ref{sec:gaugingKR} the relativistic NS-NS sector target space geometry. Specifically, we exhibit how the Kalb--Ramond field, along with the metric field, can be generated by gauging a stringy version of the Poincaré algebra, in analogy to how pseudo-Riemannian geometry can be obtained from gauging the Poincaré algebra (see for example~\cite{Andringa:2010it,Hartong:2015zia}).

In Section \ref{sec:TSNCgeometry} we show how the Nambu--Goto action for a neutral NR string with TSNC target space geometry emerges from the $c\rightarrow \infty$ limit of the action of the relativistic string in the NS-NS sector. The TSNC geometry consists of the longitudinal vielbeine $\tau^A_\mu$, the transverse metric $h_{\mu\nu}$ and the two-form $m_{\mu\nu}$.
We point out that our $c\rightarrow \infty$ limit closely parallels the point particle case, reviewed in Appendix \ref{sec:particle}. In that case when taking the $c\rightarrow \infty$ limit of an extremal relativistic particle one finds the coupling of a neutral NR particle to torsional Newton--Cartan (TNC) geometry given by the clock one-form $\tau_\mu$, the transverse metric $h_{\mu\nu}$ and the one-form  $m_\mu$.

In Section \ref{sec:symmetries_TSNC} we consider in detail the symmetries of the TSNC geometry. This involves a novel
type of string Bargmann algebra, which we call the F-string Galilei (FSG) algebra, and we show how  TSNC geometry emerges from its gauging. As part of this, we explain that the torsion of TSNC geometry is unconstrained for the classical NR string. Further, we show how the FSG algebra arises from an İnönü--Wigner contraction of the string Poincaré algebra. This is again in close analogy to  the particle case reviewed in Appendix \ref{sec:particle}. The relation between the SNC geometry and algebra \cite{Andringa:2012uz,Bergshoeff:2018yvt,Bergshoeff:2019pij} and the TSNC geometry and FSG algebra is discussed in more detail in Appendix \ref{app:algebras}.

\subsection{Kalb--Ramond field from string Poincaré symmetries}
\label{sec:gaugingKR}

In this section we show how the geometric fields $g_{\mu \nu} $ and $B_{\mu \nu}$ of the NS-NS sector of the fundamental relativistic string can be related to gauge fields taking value in a stringy extension of the Poincaré algebra. As part of this, we show how their transformation properties follow from the adjoint action in the corresponding Lie group.
For this we will use and extend results in Appendix B of \cite{Harmark:2019upf}, which was motived in part by the double field theory approach to string geometry \cite{Hull:2009mi,Hohm:2010pp}.

Our starting point consists of  the usual transformations of the metric and Kalb--Ramond field under  diffeomorphisms $\xi^\mu$ and one-form gauge transformations $\lambda_\mu$,
\begin{equation}
\label{gBtrafo}
\bar \delta g_{\mu \nu} = {\cal{L}}_\xi g_{\mu \nu} \spa \bar \delta  B_{\mu \nu} = {\cal{L}}_\xi  B_{\mu \nu} +
2 \partial_{[\mu } \lambda_{\nu]}.
\end{equation}
Following conventions in the literature, we use the notation $\bar\delta$ to distinguish these transformations from the adjoint transformations induced by the gauged algebra in question, which we denote by $\delta$ below.

It is known \cite{Neeman:1978zvv,DAuria:1982uck} that for pseudo-Riemannian geometry, it is possible to obtain the transformation of the metric from a gauging procedure of the Poincaré group.
To see this, we introduce the vielbein decomposition of the metric,
\begin{equation}
  \label{gdef}
  g_{\mu \nu} = \eta_{\ab\bb}\, e^\ab_{\mu}  e^\bb_{\nu}.
\end{equation}
Here, we have introduced frame indices $\ab,\bb,\ldots$ which are $D$-dimensional.
The vielbeine~$e^\ab_\mu$ correspond to the gauge field of translations $P_\ab$.
Likewise, the spin connection~$\omega_\mu{}^{\ab\bb}$ is the gauge field associated to Lorentz transformations $M_{\ab\bb}$.
Recall that, naively, the $D^2$ degrees of freedom in the vielbeine $e^\ab_\mu$ overparametrize the $D(D+1)/2$ degrees of freedom in the metric $g_{\mu\nu}$, but this is compensated by the invariance of the parametrization~\eqref{gdef} under the $D(D-1)/2$ local Lorentz transformations.

Similarly, to find an underlying gauge structure for the $B$-field it is  natural to introduce an additional set of generators $Q_\ab$ that transform as a vector under Lorentz transformations, with corresponding gauge fields $\pi_\mu^\ab$. We will
refer to these as the $\pi$-gauge fields.%
\footnote{This construction closely resembles the two-vielbein formalism used in double field theory \cite{Hull:2009mi,Hohm:2010pp}. It would be interesting to understand  the relation to double field theory in more detail.
See also Refs.~\cite{Morand:2017fnv,Blair:2019qwi,Gallegos:2020egk,Blair:2020gng} for work on non-relativistic geometry
and non-relativistic string theory using the double field theory formalism.}
This allows us to parametrize~\cite{Harmark:2019upf}
\begin{equation}
\label{Bdef}
  B_{\mu\nu} = \eta_{\ab\bb}\, e^\ab_{[\mu} \pi^\bb_{\nu]}.
\end{equation}
It is easy to see that this parametrization indeed captures the correct number of degrees of freedom of an antisymmetric two-tensor, since it is invariant under shifts $\pi^\ab_\mu \to \pi^\ab_\mu + S^\ab{}_\bb e^\bb_\mu$ for symmetric matrices $S_{\ab\bb}=S_{(\ab\bb)}$.
This reduces the $D^2$ degrees of freedom in $\pi^\ab_\mu$ to the required $D(D-1)/2$ of the Kalb--Ramond field.

This motivates us to consider what we will call the \emph{string Poincaré algebra},%
\begin{subequations}
  \label{eq:poincare-plus-vector-commutation-relations}
  \begin{align}
    [M_{\ab\bb}, M_{\cb\db}]
    &= \eta_{\ab\cb} M_{\bb\db}
    - \eta_{\bb\cb} M_{\ab\db}
    + \eta_{\bb\db} M_{\ab\cb}
    - \eta_{\ab\db} M_{\bb\cb},
    \\
    [M_{\ab\bb}, P_\cb]
    &= \eta_{\ab\cb} P_\bb - \eta_{\bb\cb} P_\ab,
    \\
    [M_{\ab\bb}, Q_\cb]
    &= \eta_{\ab\cb} Q_\bb - \eta_{\bb\cb} Q_\ab,
  \end{align}
\end{subequations}
consisting of the Poincaré algebra with translations $P_\ab$ and Lorentz transformations
$M_{\ab\bb}$, augmented by a second set of vectors $Q_\ab$.%
\footnote{One could also include further generators corresponding to the symmetric $S$ symmetries in the shift $ \pi \rightarrow \pi + S e$.  However, since we choose to only consider the combination \eqref{Bdef} which is invariant under these transformations, these will not be relevant in the following.}
The connection on this algebra
\begin{equation}
  \mA_\mu = e^\ab_\mu P_\ab + \frac{1}{2} \omega_\mu{}^{\ab\bb} M_{\ab\bb} + \pi^\ab_\mu Q_\ab,
\end{equation}
contains both the vielbeine $e^\ab_\mu$ and the spin connection $\omega_\mu{}^{\underline{ab}}$, as well as the novel gauge field $\pi^\ab_\mu$.
If we parametrize its local symmetry transformations by
\begin{equation}
  \Lambda
  = \zeta^\ab P_\ab + \frac{1}{2} \sigma^{\ab\bb} M_{\ab\bb} + \kappa^\ab Q_\ab,
\end{equation}
then its adjoint transformations and curvature are given by
\begin{align}
  \delta_\Lambda A_\mu
  &= \pd_\mu \Lambda + [A,\Lambda]_\mu
  = \delta e^\ab_\mu P_\ab + \frac{1}{2} \delta \omega_\mu{}^{\ab\bb} M_{\ab\bb}
  + \delta \pi^\ab_\mu Q_\ab,
  \\
  F
  &= dA + A \wedge A
  = R(P)^\ab P_\ab + \frac{1}{2} R(M)^{\ab\bb} M_{\ab\bb}
  + R(q)^\ab Q_\ab.
\end{align}
In components, this corresponds to
\begin{subequations}
  \label{eq:poincare-plus-vector-delta-transformations-and-curvatures}
  \begin{align}
    \delta e^\ab_\mu
    &= \pd_\mu \zeta^\ab
    - \omega_\mu{}^\ab{}_\bb \zeta^\bb
    + \sigma^\ab{}_\bb e^\bb_\mu,
    &
    R(P)^\ab
    &= de^\ab - \omega^\ab{}_\bb \wedge e^\bb,
    \\
    \delta \omega_\mu{}^{\ab\bb}
    &= \pd_\mu \sigma^{\ab\bb}
    - \omega_\mu{}^\ab{}_\cb \sigma^{\cb\bb}
    - \omega_\mu{}^\bb{}_\cb \sigma^{\ab\cb},
    &
    R(M)^{\ab\bb}
    &= d\omega^{\ab\bb} - \omega^\ab{}_\cb \wedge \omega^{\cb\bb},
    \\
    \delta \pi_\mu^\ab
    &= \pd_\mu \kappa^\ab
    - \omega_\mu{}^\ab{}_\bb \kappa^\bb
    + \sigma^\ab{}_\bb \pi_\mu^\bb,
    &
    R(q)^\ab
    &= d\pi^\ab - \omega^\ab{}_\bb \wedge \pi^\bb.
  \end{align}
\end{subequations}
We can now pose the question whether we can relate the $\bar\delta$ transformations~\eqref{gBtrafo} to the adjoint $\delta$ action in the string Poincaré algebra above.

Gauging the Lorentz transformations ensures local Lorentz symmetry, as required by the Einstein equivalence principle.
Diffeomorphisms can be recovered from the gauging of translations, by imposing what are known as curvature constraints, but this is only possible if the geometry is required to have zero torsion $R(P)^\ab=0$, which is often undesired in connection to non-relativistic limits and expansions.
However, there is a subsequent procedure~\cite{Hartong:2015zia} to define transformations~$\bar \delta$  involving the adjoint action $\delta_\Lambda$  in the Lie group, such that this is achieved.
It results in
\begin{subequations}
  \label{eq:poincare-delta-bar}
  \begin{align}
    \bar\delta e^\ab_\mu
    &= \delta e^\ab_\mu - \xi^\nu R(P)^\ab_{\mu\nu}
    = \LL_\xi e^\ab_\mu + \lambda^\ab{}_\bb e^\bb_\mu,
    \\
    \bar\delta \omega_\mu{}^\ab{}_\bb
    &= \delta \omega_\mu{}^\ab{}_\bb - \xi^\nu R(M)_{\mu\nu}{}^\ab{}_\bb
    = \LL_\xi \omega_\mu{}^\ab{}_\bb
    + \pd_\mu \lambda^\ab{}_\bb
    - \omega_\mu{}^\ab{}_\cb \lambda^\cb{}_\bb
    - \omega_\mu{}_{\bb\cb} \lambda^{\ab\cb}.
  \end{align}
\end{subequations}
Here, the diffeomorphisms are parametrized by $\xi^\mu = \theta^\mu_\ab\zeta^\ab$, where $\theta^\mu_\ab$ are the inverse vielbeine to $e^\ab_\mu$, and the local Lorentz transformations are parametrized by $\lambda_{\ab\bb} = \sigma_{\ab\bb} - \xi^\mu\omega_{\mu\ab\bb}$.

We now introduce a similar procedure that allows us to associate the one-form gauge transformations $\lambda_\mu$ of the Kalb--Ramond field in~\eqref{eq:poincare-delta-bar} to the transformations associated to the vector extension $Q_\ab$ of the Poincaré group.
For this, we define
\begin{align}
  \bar\delta \pi_\mu^\ab
  &= \delta \pi^\ab_\mu
  - \xi^\nu R(q)^\ab_{\mu\nu} + \frac{1}{2} k_\bb \theta^{\ab\nu} R(P)_{\mu\nu}^\bb
  \\
  &= \LL_\xi \pi_\mu^\ab
  + \lambda^\ab{}_\bb \pi^\bb_\mu
  + \pd_\mu k^\ab - \omega_\mu{}^\ab{}_\bb k^\bb
  + \frac{1}{2} k_\bb \theta^{\ab\nu} R(P)^\bb_{\mu\nu}.
\end{align}
Here, we have introduced $k^\ab = \kappa^\ab - \xi^\nu\pi_\nu^\ab$.
Using the parametrization~\eqref{eq:gauging-fsnc-boost-deltabar-transformations}, we see that the Kalb--Ramond field transforms as
\begin{align}
  \bar\delta_k B_{\mu\nu}
  &= \frac{1}{2} \eta_{\ab\bb} e^\ab_\mu
  \left(\pd_\nu k^\bb - \omega_\nu{}^\bb{}_\cb k^\cb\right)
  + \frac{1}{4} \eta_{\ab\bb} e^\ab_\mu k_\cb \theta^{\bb\rho} R(P)^\cb_{\nu\rho}
  - \left(\mu \leftrightarrow \nu\right)
  \\
  &= - \frac{1}{2} \pd_\mu \left(\eta_{\ab\bb} e^\ab_\nu k^\bb\right)
  - \frac{1}{2} k_\ab \left(\pd_\nu e^\ab_\mu - \omega_\nu{}^\ab{}_\bb e^\bb_\mu\right)
  + \frac{1}{4} k_\ab R(P)^\ab_{\nu\mu}
  - \left(\mu \leftrightarrow \nu\right)
  \\
  &= \pd_\mu \lambda_\nu - \pd_\nu \lambda_\mu,
\end{align}
 where $\lambda_\mu = - \frac{1}{2}k_\ab e^\ab_\mu$.
 This has the desired form \eqref{gBtrafo} of the one-form gauge transformation
 of the $B$-field.

With this prescription, gauging the string Poincaré algebra~\eqref{eq:poincare-plus-vector-commutation-relations} leads to the geometry of the NS-NS sector of the fundamental relativistic string (modulo the dilaton).
In particular, while the local translations $P_\ab$ are recast into diffeomorphism, the additional generators $Q_\ab$ are recast into one-form gauge transformations of the Kalb--Ramond field.
Note that this is done without any constraint on the torsion $R(P)^\ab_{\mu\nu}$ in the geometry.

As we will shortly see, the target space geometry of NR string theory along with its symmetries
can be obtained in a similar way, originating from the gauging of a certain İnönü--Wigner contraction of the string Poincaré algebra discussed above.

\subsection{NR string action with TSNC target space from \texorpdfstring{$c\rightarrow\infty$}{c->infty} limit}
\label{sec:TSNCgeometry}

Using the new perspective on the Kalb--Ramond field obtained above, we now revisit the $c\rightarrow\infty$ limit of the relativistic fundamental string in the NS-NS sector.
This will be done in close analogy with the $c\rightarrow\infty$ limit of the action of an extremal relativistic particle, which, as reviewed in Appendix~\ref{sec:particle}, leads to an uncharged non-relativistic particle in TNC geometry.
Similarly, the limiting procedure below will exhibit the importance of the fact that the fundamental relativistic string is extremal, {\sl i.e.}~that its charge with respect to the Kalb--Ramond field is equal to the string tension.
As a result we will find a novel formulation of the general target space geometry of non-relativistic strings, its symmetries, and the underlying group-theoretic structure.

Our starting point is the following action for the relativistic fundamental string in the NS-NS sector,
\begin{equation}
S=S_{\rm NG}+S_{\rm WZ},
\end{equation}
consisting of the sum of the Nambu-Goto and Wess-Zumino term, which are respectively
\begin{equation}
S_{\rm NG} = - c\, T_{\rm F} \int d^2\sigma \sqrt{-\det g_{\alpha\beta}},
\qquad
S_{\rm WZ} =  -c \frac{T_{\rm F}}{2} \int d^2\sigma B_{\alpha\beta} \epsilon^{\alpha\beta}.
\end{equation}
Here, the induced metric and Kalb--Ramond field on the string worldvolume are
\begin{equation}
g_{\alpha\beta} = \partial_\alpha X^\mu \partial_\beta X^\nu g_{\mu\nu} \spa B_{\alpha\beta} = \partial_\alpha X^\mu \partial_\beta X^\nu B_{\mu\nu},
\end{equation}
we have $\epsilon^{01}=-\epsilon_{01}$
and $T_{\rm F}$ is the string tension.
We will now use the parametrization of these relativistic NS-NS fields in terms of vielbeine $e^{\ab}_\mu$ and  $\pi^{\ab}_\mu$ given in \eqref{gdef} and \eqref{Bdef},
\begin{equation}
\label{gBdef}
g_{\mu\nu} = \eta_{\ab\bb} e^{\ab}_\mu e^{\bb}_\nu
\spa
B_{\mu\nu} = \eta_{\ab\bb} e^{\ab}_{[\mu} \pi^{\bb}_{\nu]},
\end{equation}
where $\mu = 0, \ldots , D-1$ label the $D$-dimensional spacetime coordinates and $\ab = 0, \ldots , D-1$ are frame indices.
We now decompose the frame indices $\ab=(A,a)$ corresponding to directions longitudinal and transverse to the string worldsheet, where $A=0,1$ are longitudinal and $a=2,...,d-1$ are transverse.%
\footnote{This split is also done in the derivation of SNC geometry \cite{Bergshoeff:2019pij}, but the distinction in our analysis is that we use the parametrization of the $B$-field in \eqref{gBdef}, which will be essential for the final result.}
Correspondingly, we denote
\begin{equation}
  \label{EPidecomposition}
  e^{\ab}_\mu
  = ( c E^A_\mu , e^a_\mu ) \spa
  \pi^{\ab}_\mu = ( c \Pi^A_\mu , \pi^a_\mu ),
\end{equation}
where we have introduced an explicit factor of $c$ in the longitudinal directions, so that the light cone opens up in
the transverse directions when $c\rightarrow \infty$.
With this, we find
\begin{subequations}
  \begin{align}
    g_{\alpha\beta}
    &= c^2 \eta_{AB} E^A_\alpha E^B_\beta + \delta_{ab} e^a_\alpha e^b_\beta,
    \\
    B_{\alpha\beta}
    &= c^2 \eta_{AB} E^A_{[\alpha} \Pi^B_{\beta]} + \delta_{ab} e^a_{[\alpha} \pi^b_{\beta]}.
  \end{align}
\end{subequations}

Following Equation~\eqref{TAdef}, we now reparametrize the longitudinal components $E^A$ and $\Pi^A$ of the vielbein and $\pi$-gauge field as follows,%
\footnote{We have chosen here the notation $\pi^A_\mu $ for the subleading fields to recall that they are connected to
the gauge fields entering the $B$-field. They will turn out to be related to the $m^A_\mu$ fields of the SNC geometry (see {\sl e.g.}~\cite{Bergshoeff:2018yvt}), but with some notable differences in the way they transform.
}
\begin{subequations}
\label{EPidef}
\begin{align}
  E^A_\mu
  &= \tau^A_\mu + \frac{1}{2c^2} \pi^B_\mu \epsilon_B{}^A,
  \\
  \Pi^A_\mu
  &= \epsilon^A{}_B \tau^B_\mu + \frac{1}{2c^2} \pi^A_\mu,
\end{align}
\end{subequations}
where $\epsilon_B{}^A  = \epsilon_{BC} \eta^{CA} $.
In Section \ref{sec:symmetries_TSNC} below we will see what this means at the level of the algebra, and in particular we will show that, after the $c\rightarrow \infty$ limit, this corresponds to a particular İnönü--Wigner contraction of the string Poincaré algebra given in Equation~\eqref{eq:poincare-plus-vector-commutation-relations}.

After some algebra (see also \cite{Harmark:2019upf,Bergshoeff:2019pij}), we find that with the field redefinitions~\eqref{EPidef} the integrands of the NG and WZ action can be written as
\begin{subequations}
  \label{NGWZexp}
  \begin{gather}
    \label{ter1}
    \sqrt{-\det g_{\alpha\beta}}
    = c^2 \sqrt{-\tau} \left(
      1 + \frac{1}{2c^2} \eta^{AB} \tau_A^\alpha \tau_B^\beta h_{\alpha\beta}
    \right)
    + \frac{1}{2} \epsilon^{\alpha\beta} \eta_{AB} \tau^A_\alpha \pi^B_\beta + \CO (c^{-2} ),
    \\
    \label{ter2}
    \frac{1}{2} \epsilon^{\alpha\beta} B_{\alpha\beta}
    = - c^2 \sqrt{-\tau} + \frac{1}{2} \epsilon^{\alpha\beta} B^\nparallel_{\alpha\beta}
    + \CO (c^{-2} ),
  \end{gather}
\end{subequations}
where the worldsheet induced metric $\tau_{\alpha\beta}$ corresponding to the vielbeine $\tau_\alpha^A$ and their worldsheet inverse $\tau^\alpha_A$ are given by
\begin{equation}
  \label{dettau}
  \tau_{\alpha\beta}
  = \eta_{AB} \tau^A_\alpha \tau^B_\beta
  \spa
  \tau_A^\alpha
  = - \frac{\epsilon^{\alpha\beta}\epsilon_{AB} \tau_\beta^B}{\sqrt{-\tau}},
\end{equation}
and we defined the determinant of $\tau_{\alpha\beta}$ as $\tau = \det{\tau_{\alpha\beta}}$.
Additionally, in Equation~\eqref{NGWZexp} we have used the transverse metric $h_{\mu\nu}$ and the non-longitudinal part
$B^\nparallel_{\mu\nu}$ of the Kalb--Ramond field, which are defined by
\begin{equation}
  \label{hbdef}
  h_{\mu \nu}
  = e^a_\alpha e^b_\beta \delta_{ab}
  \spa B^\nparallel_{\mu \nu} = e^{a}_{[\mu} \pi^{b}_{\nu]} \delta_{ab},
\end{equation}
satisfying $\tau_A^\mu h_{\mu \nu} =0$ and  $\tau_A^\mu \tau_B^\nu B^\nparallel_{\mu \nu} =0$, respectively. The expressions
in \eqref{NGWZexp} involve the pullbacks of  $\tau^A$, $h$ and $B^\nparallel$ on the worldvolume, so e.g.
 $\tau^A_\alpha = \partial_\alpha X^\mu \tau^A_\mu$ etc.
Adding the two terms \eqref{ter1}, \eqref{ter2} we see that the divergent leading-order terms cancel with each other.
The reason one can accomplish this is that the fundamental string is extremal.
This is analogous to the cancellation seen for an extremal particle (see Appendix~\ref{sec:particle}).

After rescaling $c\, T_{\rm F}=T$, taking $c\to\infty$ then leads%
\footnote{Note that we are taking $\tau^A_\mu$ and $\pi^A_\mu$ to be $\OO(c^0)$ and, in a slight abuse of notation, we are denoting their leading-order contributions by the same symbol.}
to the (Nambu-Goto form of the) non-relativistic fundamental string action,
\begin{equation}
  \label{NRaction2}
  S_{\rm NR}
  =  -\frac{T}{2} \int d^2 \sigma \left[
    \sqrt{-\tau} \, \eta^{AB} \tau_A^\alpha \tau_B^\beta h_{\alpha\beta} +
    \epsilon^{\alpha\beta} m_{\alpha\beta}
  \right],
\end{equation}
where we have defined the two-form $m_{\mu\nu}$ as
\begin{equation}
  \label{eq:mdef}
  m_{\mu \nu}
  = \eta_{AB} \tau^A_{[\mu} \pi^B_{\nu]} + \delta_{ab} e^a_{[\mu} \pi^b_{\nu]}.
\end{equation}
We emphasize that the form \eqref{NRaction2} can be regarded as the direct analogue of the coupling of a neutral non-relativistic particle to TNC geometry, which only couples to fields in the geometry and no other electromagnetic-type fields.
While the TNC geometry is parametrized by $\tau_\mu$, $h_{\mu \nu}$ and $m_\mu$, we see from \eqref{NRaction2} that the non-relativistic string couples to the geometry
\begin{equation}
  \label{Fstringgeometry}
  \mbox{torsional string Newton--Cartan geometry}:
  \qquad
  \tau_\mu^A,
  \quad
  h_{\mu \nu},
  \quad
  m_{\mu \nu}.
\end{equation}
We will see from the transformation rules discussed in the next section that the TSNC two-form $m_{\mu \nu}$ should be considered as an intrinsic part of the string geometry, in the same way that TNC geometry contains the one-form $m_\mu$ as one of the geometrical variables.
In that sense, one should view the original Kalb--Ramond coupling as having disappeared in the limit, just as the extremal relativistic particle becomes a neutral massive non-relativistic particle in the NR limit.
This is the reason for our choice of notation $m_{\mu \nu}$.

We also note that for a string that is point-like  $X^{\hat\mu} (\tau)$ in all directions $\hat\mu = 0 \ldots D-2 $  except for the direction $X^v = R \sigma$
in which there is a non-zero winding, we see that  string action~\eqref{NRaction2} reduces to the action of a NR particle with mass $2\pi RT$, where the NC gauge field $m_{\hat\mu}$ is given by $\int d \sigma \, m_{\hat\mu v}$.
Furthermore, just like the variation of $m_{\hat\mu}$ in the worldline action of the NR particle returns the mass current $T^{\hat\mu} = m\int d\tau\,  \partial_\tau X^{\hat\mu} \delta (x - X(\tau)) $,  the response to varying $ m_{\mu \nu}$ corresponds to the {\sl worldsheet tension current}
\begin{equation}
  J_{\rm T}^{\mu \nu}
  = T \int d^2\sigma\, \epsilon^{\alpha \beta} \partial_\alpha X^{\mu} \partial_\beta X^{\nu}  \delta( x - X(\sigma^\alpha)),
\end{equation}
of the NR string, which is conserved $\partial_\mu J_{\rm T}^{\mu \nu}=0$ as a consequence
of the gauge invariance of the NC string potential.

We can compute the corresponding conserved charges of $J_{\rm T}^{\mu \nu}$ as $M^\mu = \int d \sigma J^{0 \mu}$.
We compute these charges choosing the world-volume coordinates as $\tau= X^0$ and $\sigma = X^v$, with $X^{v}$ the compact longitudinal direction with periodicity $ 2\pi R$.
Assuming that the string is not compactified in any of the transverse directions we see that the only non-zero charge is  $M^v = 2 \pi T R $.
This charge can be considered to be the gravitational mass under the natural interpretation that $m_{0v} $ is the Newtonian potential by looking at the second term in the action~\eqref{NRaction2}.
Finally, we note that since both the first (kinetic) term and the second (gravitational interaction) term in
the action \eqref{NRaction2} come with the same coupling, it appears that non-relativistic strings satisfy a version of
the weak equivalence principle.

All this shows that the torsional string Newton--Cartan (TSNC) geometry \eqref{Fstringgeometry} can be regarded as the string counterpart of  TNC geometry for particles \cite{Christensen:2013lma,Christensen:2013rfa,Hartong:2015wxa}.

In the next section we will discuss the transformation of the fields in \eqref{Fstringgeometry} and the corresponding invariance of the NR string action \eqref{NRaction2}. We will also find the underlying symmetry algebra of TSNC geometry, which corresponds to an İnönü--Wigner contraction of the string Poincaré algebra.

Before that, a couple of remarks are in order.
First of all, as we shall see in Section~\ref{sec:null_reduction}, the action~\eqref{NRaction2} agrees with the action one
obtains via null reduction of the relativistic fundamental string in the NS-NS sector, as originally derived in~\cite{Harmark:2019upf}.
This is satisfying since the coupling of an NR particle to TNC geometry can be obtained both via a $c\rightarrow \infty$ limit as well as via null reduction, as reviewed in Appendix \ref{sec:particle}.
In Section \ref{sec:null_reduction} we will show how this can be achieved from the null reduction point of view and we will find agreement with the results presented in this section.

Another comment is that, using the expression~\eqref{dettau} for $\tau^\alpha_A$ in terms of the matrix of cofactors, the action \eqref{NRaction2} can be rewritten as
\begin{equation}
  \label{NRactionSNC}
  S_{\rm NR} =  -\frac{T}{2} \int d^2 \sigma \left[ \sqrt{-\tau} \, \eta^{AB} \tau_A^\alpha \tau_B^\beta \bar h_{\alpha\beta} +
  \epsilon^{\alpha\beta} B^\nparallel_{\alpha\beta} \right],
\end{equation}
where $B^\nparallel_{\mu \nu}$ is the non-longitudinal part of $m_{\mu\nu}$, and
\begin{equation}
  \bar h_{\mu \nu}
  =  h_{\mu \nu} - \frac{1}{2} \epsilon_{AB} ( \tau_\mu^A \pi_\nu^B +  \tau_\nu^A \pi_\mu^B ).
\end{equation}
Upon identifying $\pi^A$ with the SNC field $m^A_\mu$ via
$\pi^A_\mu = - 2\epsilon^A{}_B m^B_\mu$,
this form is closely related to the SNC action \eqref{SNC_action}, though with
the difference that in the SNC action the non-longitudinal field $B^\nparallel_{\mu\nu}$ is replaced by an unconstrained two-form field.
This clarifies (see also~\cite{Harmark:2019upf})
the origin of the Stückelberg symmetry in the SNC action, which is a redundancy that transforms the longitudinal part of $m_{\mu\nu}$ into the
$ 2\eta_{AB} \tau_{(\mu}^A m_{\nu)}^B$ part of $\bar h_{\mu \nu}$.

\subsection{Symmetries of torsional string Newton--Cartan geometry}
\label{sec:symmetries_TSNC}

We now turn to the local symmetries of the action \eqref{NRaction2} and the underlying non-relativistic space-time symmetry algebra of the TSNC target space geometry.
From Section~\ref{sec:gaugingKR}, we know how the symmetries of the NS-NS fields $g_{\mu\nu}$ and $B_{\mu\nu}$ can be derived from the string Poincaré algebra~\eqref{eq:poincare-plus-vector-commutation-relations} in terms of the transformations of relativistic vielbein $E^{\bar{a}}_\mu$ and $\pi$-gauge field $\Pi^{\bar{a}}_\mu$.
Building on the field redefinition~\eqref{EPidef} that led to the non-relativistic string action~\eqref{NRaction2} in the $c\to\infty$ limit, we can then define a contraction of the string Poincaré algebra.
Following a similar procedure as in the relativistic setting of Section~\ref{sec:gaugingKR}, the gauging of the resulting algebra then allows us to reproduce all local symmetries of the background TSNC geometry.
This provides a concise definition of the geometry and is among one of the central results of the paper.

For simplicity, we first present the resulting transformations and discuss the underlying algebra afterwards.
First of all, there are local $SO(1,1)$ boost symmetries rotating the longitudinal  fields $(\tau^{A}_\mu,\pi^{A}_\mu)$ and local $SO(D-2)$  rotations acting on the transverse fields $(e^{a}_\mu,\pi^{a}_\mu)$.
In addition, the action \eqref{NRaction2} is manifestly invariant under the F-string Newton--Cartan one-form gauge transformation,
\begin{equation}
\label{2formtrafo}
\bar \delta m_{\mu \nu} = 2\partial_{[\mu} \lambda_{\nu]}.
\end{equation}
Finally, the action is invariant under string boosts parameterized by $\lambda^{Ab}$, which act on the fundamental fields as
\begin{subequations}
  \label{eq:gauging-fsnc-boost-deltabar-transformations}
  \begin{align}
    \bar\delta\tau^A_\mu
    &= 0,
    \\
    \bar\delta e^a_\mu
    &= - \lambda_B{}^a \tau^B_\mu,
    \\
    \bar\delta \pi^A_\mu
    &= \lambda^A{}_b \pi^b_\mu - \epsilon^A{}_B \lambda^B{}_c e^c_\mu,
    \\
    \bar\delta \pi^a_\mu
    &= - \epsilon_{BC} \lambda^{Ba} \tau^C_\mu.
  \end{align}
\end{subequations}
From the definition of $h_{\mu \nu}$ in \eqref{hbdef} and $m_{\mu \nu}$  in \eqref{eq:mdef} it is easy to see that these composite fields then transform as
\begin{equation}
\label{eq:tsnc-boost-composite-fields}
    \bar\delta h_{\mu\nu}
    = - \lambda_{Ab} \left(
      \tau^A_\mu e^b_\nu + \tau^A_\nu e^b_\mu
    \right)
    \spa
    \bar\delta m_{\mu\nu}
    = - 2 \epsilon_{AB} \lambda^B{}_c \tau^A_{[\mu} e^c_{\nu]}.
\end{equation}
Using the second identity~\eqref{dettau} it follows that these local string Galilean boost transformations indeed leave the action~\eqref{NRaction2} invariant.

In Section~\ref{sec:TSNCgeometry}, we have seen that the NR string naturally couples to TSNC geometry, and in the above we have outlined its transformations.
We now derive the corresponding algebra that results from the $c\to\infty$ contraction of the string Poincaré algebra~\eqref{eq:poincare-plus-vector-commutation-relations}.
Next, following the procedure introduced in Section~\ref{sec:gaugingKR}, we outline how the transformations discussed above follow from a gauging of the resulting algebra.

First of all, in accordance with \eqref{EPidecomposition} we decompose the Poincaré generators $P_\ab$ and $M_{\ab\bb}$ using the split $\ab = (A,a)$ in longitudinal and transverse indices.
We denote the resulting generators by
\begin{equation}
  P_A,
  \quad
  Q_A,
  \quad
  P_a,
  \quad
  Q_a,
  \quad
  J_{AB}=\epsilon_{AB}J= M_{AB},
  \quad
  J_{ab}=M_{ab},
  \quad
  c\,G_{Ab}=M_{Ab}.
\end{equation}
Next we observe that the crucial field redefinitions~\eqref{EPidef} on the longitudinal vielbeine imply at the level of the algebra the following $c$-dependent basis transformation for the generators $P_A$ and $Q_A$,
\begin{equation}
  H_A = c (P_A +  Q_B \epsilon^B{}_A) \quad,  \quad
  N_A = \frac{1}{2c} (\epsilon_A{}^B  P_B +  Q_A)
\end{equation}
Note the analogy with the combination \eqref{HNcombo} when considering the contraction of the Poincaré times $U(1)$ algebra in the particle case.
Here we obtain a similar redefinition involving the worldsheet Hamiltonian, the worldsheet translation operator and the two charges that generate the longitudinal $B$-field gauge transformations.
In the $c\to\infty$ limit, this leads to the commutation relations
\begin{subequations}
  \label{eq:fsnc-algebra-dual-rotations}
  \begin{align}
    [J_{ab}, J_{cd}]
    &= \delta_{ac} J_{bd}
    - \delta_{bc} J_{ad}
    + \delta_{bd} J_{ac}
    - \delta_{ad} J_{bc},
    \\
    [J,G_{Ab}]
    &= \epsilon^C{}_A G_{Cb},
    \\
    [J_{ab}, G_{Cd}]
    &= \delta_{ad} G_{Cb} - \delta_{bd} G_{Ca},
    \\
    [J, H_A]
    &= \epsilon^B{}_A H_B,
    \\
    [J, N_A]
    &= \epsilon^B{}_A N_B,
    \\
    [G_{Ab}, H_C]
    &= \eta_{AC} P_b + \epsilon_{AC} Q_b,
    \label{GH}
    \\
    [G_{Ab}, P_c]
    &= - \delta_{bc} \epsilon_A{}^B N_B,
    \label{GPt}
    \\
    [G_{Ab}, Q_c]
    &= - \delta_{bc} N_A,
    \label{GQ}
    \\
    [J_{ab},P_c]
    &= \delta_{ac} P_b - \delta_{bc} P_a,
    \\
    [J_{ab},Q_c]
    &= \delta_{ac} Q_b - \delta_{bc} Q_a.
  \end{align}
\end{subequations}
We call this the F-string Galilei (FSG) algebra, which has generators
$J$, $J_{ab}$ and $G_{Aa}$, corresponding to $SO(1,1)$, $SO(D-2)$ and string Galilean boosts, non-relativistic worldsheet translation operators $H_A$ and transverse translations $P_a$.
In addition, there are the generators $N_A$, $Q_a$, which originate from the NS-NS 2-form.
As we will see, they are related to the TSNC 2-form $m_{\mu \nu}$.

If one now introduces the FSG-valued connection
\begin{equation}
  \mA_\mu
  = \tau^A_\mu H_A + e^a_\mu P_a
  + \omega_\mu J + \frac{1}{2} \omega_\mu{}^{ab} J_{ab} + \omega_\mu{}^{Ab} G_{Ab}
  + \pi^A_\mu Z_A + \pi^a_\mu Q_a,
\end{equation}
one can check, using similar steps as in the discussion in Section \ref{sec:gaugingKR}, that one can reproduce the transformations \eqref{2formtrafo} and \eqref{eq:gauging-fsnc-boost-deltabar-transformations} and all other local symmetries of the action~\eqref{NRaction2} from the adjoint transformations coming from the FSG algebra.
For example, the local Galilean boosts $\lambda^{Ab}$ above are associated to the transformation parameter of the boost generator $G_{Ab}$.
Further details of this construction are given in Appendix~\ref{app:fsnc-gauging}.
Crucially, the resulting realization of the TSNC symmetries does not require any constraints on the torsion $R(H)^A_{\mu\nu}$ of the background.

It is interesting to compare this algebra to the SNC algebra (see Equations~\eqref{eq:app-string-galilei-algebra} and~\eqref{eq:app-snc-algebra-additional-commutators} in Appendix \ref{app:algebras}).
They have in common the commutator \eqref{GPt} if we dualize $N_A$ into $Z_A=\epsilon_A{}^B N_B$, mirroring the field redefinition $\pi^A_\alpha = - 2\epsilon^A{}_B m^B_\alpha$ that led to the alternative form~\eqref{NRactionSNC} of the string action.
However, the SNC algebra has a generator $Z_{AB}$ entering the commutator $[G_{Ab}, G_{Cd}] =  \delta_{ab} Z_{AB}$, which is zero in the algebra above, but needed in the SNC algebra to close the algebra.
The role of the extra gauge field corresponding to $Z_{AB}$ is not clear, as it does not enter the SNC action.
On the other hand, the algebra above features an extra set of generators $Q_a$ that are connected to the non-longitudinal part of $m_{\mu \nu}$.
This makes the appearance of the $N_A$ symmetry  distinctly different, as it also appears on the right-hand side in \eqref{GQ} while the commutator \eqref{GH} has an extra term involving $Q_a$.
This new perspective on the $N_A$  (or $Z_A$) symmetry shows that it should be identified as the gauge transformation on the $m$-field with the gauge parameter in the longitudinal direction.

Importantly, as we emphasized earlier, this symmetry algebra and its realization makes the $Z_A$ symmetry manifest and does not require a torsion constraint $R(H)^A_{\mu\nu}=0$, as is the case in the SNC action~\eqref{SNC_action}.
This mirrors the situation for a NR particle coupling to TNC geometry, where one does not find a torsion constraint.
Moreover, in the null reduction approach to constructing the NR string action, one does not find a torsion constraint either.
In fact, the $Z_A$ symmetry can also be understood from the null reduction point of view, as we will show in Section~\ref{sec:null_reduction} below.

Finally, the above strongly suggests that it is natural to consider the FSG algebra as the string analogue of the Bargmann algebra in the particle case, in the sense that it is obtained from a İnönü--Wigner contraction of the string Poincaré algebra.
It is interesting that, unlike the particle case, in string theory one does not need to invoke an extraneous $U(1)$ field, as the $B$-field is already present from the outset, since  the relativistic fundamental string by itself is already an extremal object, providing the necessary charge fields to take the non-relativistic limit.

%%%%%%%%%%%%%%%%%%%%%%%%%%%%%%%%%%%%%%%%%%%%%%%%%%%%%%%%%%%
\section{Torsional string Newton--Cartan target space from null reduction}
\label{sec:null_reduction}

In the previous section we obtained the neutral NR string from a $c \rightarrow \infty$ limit of the relativistic string in the NS-NS sector. In this section, we show that one can also obtain the same neutral NR string from a null reduction of the relativistic string in the NS-NS sector. This is analogous to the null reduction in the particle case (see Appendix \ref{sec:particle}).

%%%%%%%%%%%%%%%%%%%%%%%%%%%%%%%%%%%%%%%%%%%%%%%%%%%%%%%%%%%
\subsection{Direct derivation of TSNC action from null reduction}
\label{sec:nullred_action}

Consider the null reduction of the NS-NS sector of string theory where $\partial_u$ is a null Killing vector field. Write the metric of the null-reduced background as
\begin{equation}
\label{nullredmetric}
ds^2 =g_{MN} dx^{M}dx^{N}
= 2\tau (du-m) + h_{\hat{\mu}\hat{\nu}} dx^{\hat{\mu}} dx^{\hat{\nu}} = 2 \tau du + \bar{h}_{\hat{\mu}{\hat{\nu}}} dx^{\hat{\mu}} dx^{\hat{\nu}},
\end{equation}
with $\hat{\mu},{\hat{\nu}}=0,1,...,D-2$ and where $\tau_{\hat{\mu}}$, $m_{\hat{\mu}}$ and $h_{\hat{\mu}{\hat{\nu}}}$ do not depend on $u$.
We write the Kalb--Ramond field as
\begin{equation}
B = \frac{1}{2} B_{MN} dx^{M} \wedge dx^{N} =  \frac{1}{2} B_{\hat{\mu}{\hat{\nu}}} dx^{\hat{\mu}} \wedge dx^{{\hat{\nu}}} + b_{\hat{\mu}} du \wedge dx^{\hat{\mu}},
\end{equation}
so that $B_{u\hat{\mu}}= b_{\hat{\mu}}$.
As shown in \cite{Harmark:2017rpg,Harmark:2018cdl,Harmark:2019upf}, the Lagrangian that one obtains by null-reducing the Nambu-Goto Lagrangian of the relativistic NS-NS string is
\begin{equation}
  \begin{split}
    \label{TNC_NGL}
    \CL
    &= T \Bigg[
      h_{\hat{\mu}{\hat{\nu}}} \frac{\epsilon^{\alpha\alpha'}\epsilon^{\beta\beta'} [  ( \partial_{\alpha'} \eta + b_{\alpha'} )( \partial_{\beta'} \eta + b_{\beta'} ) - \tau_{\alpha'}\tau_{\beta'} ] }{2\epsilon^{\gamma\gamma'} \tau_{\gamma} ( \partial_{\gamma'} \eta + b_{\gamma'} )}
      \\
      &{}\qquad\qquad\qquad\qquad\qquad
      - \epsilon^{\alpha\beta} m_{\hat{\mu}} ( \partial_{\hat{\nu}} \eta + b_{\hat{\nu}} )  - \frac{1}{2} \epsilon^{\alpha\beta} B_{{\hat{\mu}}{\hat{\nu}}}
    \Bigg] \partial_\alpha X^{\hat{\mu}} \partial_\beta X^{\hat{\nu}},
  \end{split}
\end{equation}
where the pullbacks $\tau_\gamma$ and $b_\gamma$ are with respect to $X^{\hat\mu}$.
Here we have introduced the extra direction $v$ belonging to that geometry, whose embedding coordinate we denote by $X^v=\eta$. The direction $v$ is an isometry and can be thought of as dual to $u$ \cite{Harmark:2018cdl}.
This is known as the TNC string Nambu-Goto Lagrangian \cite{Harmark:2017rpg,Harmark:2018cdl,Harmark:2019upf}.

We now make contact with the TSNC geometry description.
We collect the indices of the total target space as $\mu=(\hat{\mu},v)$. The embedding map of the string is thus
\begin{equation}
X^\mu = (X^{\hat{\mu}}, X^v) \quad \mbox{where} \quad X^v=\eta\,.
\end{equation}
We introduce the target space one-forms $\tau^A_{\mu}$ with $A=0,1$ as
\begin{equation}
\label{tauA}
\tau^0_\mu =( \tau_{\hat{\mu}} ,0 )
\spa
\tau^1_\mu = (b_{\hat{\mu}} , 1).
\end{equation}
With this, the pullback of $\tau^1_\mu$, with respect to $X^\mu$, is
\begin{equation}
\tau^1_\alpha = \tau^1_\mu \partial_\alpha X^\mu = b_{\hat{\mu}} \partial_\alpha X^{\hat{\mu}} + \partial_\alpha \eta = b_\alpha + \partial_\alpha \eta.
\end{equation}
We introduce the transverse target space metric $h_{\mu\nu}$, $\mu,\nu=0,1,...,D-1$, by extending $h_{{\hat{\mu}}\hat{\nu}}$ so that
\begin{equation}
h_{vv}= h_{{\hat{\mu}}v}=0.
\end{equation}
We introduce a new two-form field $m_{\mu\nu}$ whose components are
\begin{equation}
m_{\hat{\mu}\hat{\nu}} = B_{\hat{\mu}\hat{\nu}} + m_{\hat{\mu}} b_{\hat{\nu}} - b_{\hat{\mu}} m_{\hat{\nu}} \spa m_{v\hat{\mu}} = -m_{\hat{\mu}}\,.
\end{equation}
Then \eqref{TNC_NGL} becomes
\begin{equation}
\label{TNC_NGL2}
\CL= \frac{T}{2} \left[  h_{\mu\nu} \frac{\epsilon^{\alpha\alpha'}\epsilon^{\beta\beta'}
\eta_{AB} \tau^A_{\alpha'} \tau^B_{\beta'} }{\epsilon^{\gamma\gamma'} \tau^0_{\gamma} \tau^1_{\gamma'}}   - \epsilon^{\alpha\beta} m_{\mu \nu} \right] \partial_\alpha X^{\mu} \partial_\beta X^{\nu}.
\end{equation}
Using the definitions \eqref{dettau} we can rewrite this as
\begin{equation}
\label{TNC_NGL3}
\CL= \frac{T}{2} \left[ -\sqrt{-\tau}\, \eta^{AB} \tau^\alpha_A \tau^\beta_B h_{\mu\nu}    - \epsilon^{\alpha\beta} m_{\mu \nu} \right] \partial_\alpha X^{\mu} \partial_\beta X^{\nu}.
\end{equation}
Thus, we have reproduced the Nambu-Goto Lagrangian of the NR string with TSNC geometry as target space \eqref{NRaction2}. Note that this is with the restriction that $v$ is an isometry\footnote{The $c\rightarrow\infty$ perspective does not seem to tell us that $v$ must be an isometry, but this property is crucial if one wants to insist that the string is pure winding along $v$, i.e. that the total momentum in the $v$-direction is zero and conserved (see Equation (2.16) of \cite{Harmark:2019upf}). It would be interesting to investigate further whether it is at all possible to consider NR strings for which the $v$-direction is not an isometry.}. Below we check that one also obtains the correct boost transformations and one-form gauge transformations of this geometry.

Note that there is an arbitrariness in translating $h_{{\hat{\mu}}\hat{\nu}}$ and $B_{\hat{\mu}\hat{\nu}}$ to $h_{\mu\nu}$ and $m_{\mu\nu}$ since the actions \eqref{TNC_NGL2} and \eqref{TNC_NGL3} are invariant under the transformation
\begin{equation}
\label{arbi}
h_{\mu\nu}\rightarrow h_{\mu\nu} + \tau_\mu^1 V_\nu + V_\mu \tau_\mu^1
\spa
m_{\mu\nu}\rightarrow m_{\mu\nu} + V_\mu  \tau_\nu^0 - \tau_\mu^0  V_\nu,
\end{equation}
for any $V_\mu$.
Since we demand $h_{\mu\nu}$ to be a transverse metric we need to impose the constraint $v_A^\mu V_\mu = 0$ for this transformation where $v_A^\mu$ is the inverse vielbeine in the longitudinal directions, as defined in Equation~\eqref{eq:app-inverse-vielbeine-def}.
This will be important below.

%%%%%%%%%%%%%%%%%%%%%%%%%%%%%%%%%%%%%%%%%%%%%%%%%%%%%%%%%%%
\subsection{Boost transformations}
\label{sec:nullred_boost}

We check here that one obtains the correct boost transformations of TSNC geometry from null reduction. To this end, we introduce the transverse vielbeine $e^a_{\hat{\mu}}$ such that
\begin{equation}
h_{\hat{\mu}\hat{\nu}}=\delta_{ab} e^a_{\hat{\mu}} e^b_{\hat{\nu}},
\end{equation}
with flat indices labelled $a,b=2,...,D-2$. The TNC Galilean boost transformation is
\begin{equation}
\delta \tau_{\hat{\mu}} = 0  \spa \delta m_{\hat{\mu}} = \lambda_a e^a_{\hat{\mu}}  \spa \delta e^a_{\hat{\mu}} = \lambda^a  \tau_{\hat{\mu}} \spa \delta b_{\hat{\mu}} = 0 \spa \delta B_{\hat{\mu}\hat{\nu}} =0,
\end{equation}
where $\lambda_a = \delta_{ab} \lambda^b$. This gives
\begin{equation}
\delta h_{\hat{\mu}\hat{\nu}}= \lambda_a (  \tau_{\hat{\mu}} e^a_{\hat{\nu}} +  e^a_{\hat{\mu}} \tau_{\hat{\nu}} ).
\end{equation}
We can now translate this to TSNC geometry. The transverse vielbeine are
\begin{equation}
e^a_\mu = ( e^a_{\hat{\mu}} , 0 ).
\end{equation}
Hence we find that
\begin{equation}
h_{\mu\nu}=\delta_{ab} e^a_{\mu} e^b_{\nu}.
\end{equation}
The TNC Galilean boost transformation then becomes
\begin{equation}
\label{boost1}
\delta \tau^A_{\mu} = 0   \spa \delta e^a_{\mu} = \lambda^a  \tau^0_{\mu}  \spa \delta m_{\mu\nu} = \lambda_a (e^a_\mu \tau^1_\nu - \tau^1_\mu e^a_\nu).
\end{equation}

The other string Galilei boost arises from a redundancy
in the translation to TSNC variables~\eqref{arbi}. Since $v^\mu_A V_\mu=0$ we see that we need
\begin{equation}
V_\mu = \bar{\lambda}_a e^a_\mu,
\end{equation}
for some $\bar{\lambda}_a$, hence this corresponds to the transformation
\begin{equation}
\label{boost2}
\delta \tau^A_{\mu} = 0   \spa \delta e^a_{\mu} = \bar{\lambda}^a  \tau^1_{\mu}  \spa \delta m_{\mu\nu} = \bar{\lambda}_a (e^a_\mu \tau^0_\nu - \tau^0_\mu e^a_\nu).
\end{equation}
Combining \eqref{boost1} and \eqref{boost2} we find
\begin{equation}
\label{boost-total}
\delta \tau^A_{\mu} = 0   \spa \delta e^a_{\mu} = \lambda^{a} {}_A  \tau^A_{\mu}  \spa \delta m_{\mu\nu} = \lambda_a {}^A \epsilon_{AB} (e^a_\mu \tau^B_\nu - \tau^B_\mu e^a_\nu),
\end{equation}
where $\lambda_{a}{}^A = \eta^{AB} \lambda_{aB}$ and we identify $\lambda_{a0}=\lambda_a$ and $\lambda_{a1}=\bar{\lambda}_a$.  Comparing to Eqs.~\eqref{eq:gauging-fsnc-boost-deltabar-transformations} and~\eqref{eq:tsnc-boost-composite-fields},
this shows that we obtain the full boost transformation of TSNC geometry.

%%%%%%%%%%%%%%%%%%%%%%%%%%%%%%%%%%%%%%%%%%%%%%%%%%%%%%%%%%%
\subsection{One-form gauge transformations}
\label{sec:nullred_ZA}

Having established the string Galilei boost transformations of TSNC geometry, we now verify
that one also obtains the one-form gauge transformations of TSNC geometry from the null reduction
perspective.  We begin by noting that the TNC action has the $U(1)$ symmetry (see Eqs.~(2.29) and (2.30) of \cite{Harmark:2019upf})
\begin{equation}
\label{eq:u1-gauge-tr-null-red}
\delta \tau_{\hat{\mu}} = 0 \spa \delta m_{\hat{\mu}} = \partial_{\hat{\mu}} \chi \spa \delta e^a_{\hat{\mu}} = 0
\spa \delta b_{\hat{\mu}} = 0 \spa \delta B_{\hat{\mu}\hat{\nu}} = b_{\hat{\mu}} \partial_{\hat{\nu}} \chi-  b_{\hat{\nu}} \partial_{\hat{\mu}} \chi .
\end{equation}
Notice that there are no requirements on $\tau_{\hat{\mu}}$ or $b_{\hat{\mu}}$.
In terms of the TSNC variables this corresponds to
\begin{equation}
\delta \tau^A_{\mu} = 0   \spa \delta e^a_{\mu} = 0  \spa \delta m = - dv \wedge d\chi.
\end{equation}
We see this is a one-form gauge transformation of the two-form $m_{\mu\nu}$
\begin{equation}
\delta m_{\mu\nu} = (da)_{\mu\nu} \spa a_\mu  = \chi \delta^v_\mu,
\end{equation}
since $a = \chi dv$.
Thus, the TNC $U(1)$ gauge transformations in Equation~\eqref{eq:u1-gauge-tr-null-red} comprise a subset of one-form gauge transformations of $m_{\mu\nu}$.
From this it is clear that TNC $U(1)$ gauge transformations do not constrain the torsion of the TSNC geometry as they are symmetries regardless of whether $(d\tau)_{\hat{\mu}\hat{\nu}} \neq 0$ or $db_{\hat{\mu}\hat{\nu}} \neq 0$ \cite{Harmark:2019upf}.

In addition to the above transformations one can consider any of the following one-form transformations in TNC variables
\begin{equation}
\delta \tau_{\hat{\mu}} = 0 \spa \delta m_{\hat{\mu}} = 0 \spa \delta e^a_{\hat{\mu}} = 0
\spa \delta b_{\hat{\mu}} = 0 \spa \delta B_{\hat{\mu}\hat{\nu}} = \partial_{\hat{\mu}} c_{\hat{\nu}} - \partial_{\hat{\nu}} c_{\hat{\mu}}.
\end{equation}
Writing
\begin{equation}
c_\mu = ( c_{\hat{\mu}} , 0),
\end{equation}
we see that these transformations are also one-form transformations in the TSNC variables
\begin{equation}
\delta \tau^A_{\mu} = 0   \spa \delta e^a_{\mu} = 0  \spa \delta m_{\mu\nu} = \partial_\mu c_\nu - \partial_\nu c_\mu.
\end{equation}
Forming the combined transformation
\begin{equation}
\lambda_\mu = c_\mu + \chi \delta^v_\mu,
\end{equation}
we see that we have obtained all the one-form transformations of the two-form $m_{\mu\nu}$ with
\begin{equation}
\delta \tau^A_{\mu} = 0   \spa \delta e^a_{\mu} = 0  \spa \delta m_{\mu\nu} = \partial_\mu \lambda_\nu - \partial_\nu \lambda_\mu.
\end{equation}
Again, this has been obtained without having to restrict $\tau^A_\mu$.
In conclusion, we have shown that one obtains all the one-form gauge transformations of TSNC geometry.

\section{Conclusions}\label{sec:conclusion}

In this paper we have shown that the
torsional string Newton--Cartan (TSNC) geometry, defined by the fields
\begin{equation}
\label{con_FSNC}
    \tau^A_\mu, \ h_{\mu\nu}, \ m_{\mu\nu},
\end{equation}
is the natural target space of the non-relativistic (NR) string of \cite{Gomis:2000bd,Danielsson:2000gi,Harmark:2017rpg,Kluson:2018egd,Bergshoeff:2018yvt,Harmark:2018cdl,Gallegos:2019icg,Harmark:2019upf} when thought of as arising from a large $c$ limit or a null reduction. This is in analogy with the non-relativistic point particle case reviewed in Appendix \ref{sec:particle}.
In particular, the TSNC two-form $m_{\mu\nu}$ is part of the geometry of the NR string and generalizes the particle Newton--Cartan one-form $m_\mu$. The NR string is seen to be neutral and the component $m_{0v}$ can be interpreted as the Newtonian potential, coupling to the mass density two-form current of the NR string.

The corresponding Nambu--Goto action for the string is
\begin{equation}
\label{NRNG}
S_{\rm NR,NG} =  -\frac{T}{2} \int d^2 \sigma \left[ \sqrt{-\tau} \eta^{AB} \tau_A^\alpha \tau_B^\beta h_{\alpha\beta} +
 \epsilon^{\alpha\beta} m_{\alpha\beta} \right] + \frac{1}{4\pi} \int d^2 \sigma \,\sqrt{-\tau} R^{(2)} \phi\,,
\end{equation}
where, for completeness, we have added the dilaton term in comparison to \eqref{NRaction2}. It is straightforward to write down the corresponding Polyakov action,
\begin{align}
\label{BGY}
S_{\rm{NR,Pol}}
&= - \frac{T}{2}\Big[ \sqrt{-\gamma}\, \gamma^{\alpha\beta}\partial_\alpha X^\mu \partial_\beta X^\nu h_{\mu \nu} +\eps^{\alpha\beta}\partial_\alpha X^\mu \partial_\beta X^\nu m_{\mu \nu}
\\ \nonumber
&{}\qquad\qquad\qquad
+ \lambda \eps^{\alpha\beta}  e_\alpha{}^+ \tau_\mu{}^+ \partial_\beta X^\mu + \bar\lambda\eps^{\alpha\beta}  e_\alpha{}^- \tau_\mu{}^- \partial_\beta X^\mu  \Big]
+\frac{1}{4\pi}e {\cal{R}}^{(2)}\Phi\, ,
\end{align}
where $\tau_\mu^{\pm} =\tau_\mu^0 \pm \tau_\mu^1$.
Here, we have also expressed the
worldsheet metric in terms of zweibeine
$\gamma_{\alpha\beta}=e_\alpha{}^a e_\beta{}^b \eta_{ab}$, and defined  $e_\alpha^\pm = e_\alpha^0 \pm e_\alpha^1$.
The TSNC geometry given by the fields \eqref{con_FSNC} can be obtained by gauging the novel F-string Galilei (FSG) algebra given in \eqref{eq:fsnc-algebra-dual-rotations}.
The salient transformations of the TSNC geometry are written in \eqref{2formtrafo} and \eqref{eq:tsnc-boost-composite-fields}, which include the string Galilean boosts
as well as the one-form gauge transformations of the TSNC two-form $m_{\mu\nu}$.  The  action \eqref{NRNG} is invariant under these transformations and, in particular, the one-form gauge symmetry replaces the $Z_A$ symmetry of \cite{Bergshoeff:2018yvt} in a manner that removes the necessity of enforcing a foliation constraint.

We have shown that the Nambu--Goto action \eqref{NRNG} arises from an infinite speed of light limit of the relativistic string in the NS-NS sector. With this, the fields \eqref{con_FSNC} and their transformations
\eqref{eq:tsnc-boost-composite-fields} and \eqref{2formtrafo} emerge naturally. To accomplish this, we have reinterpreted the target space of the relativistic string as originating from the  gauging of what we call a string Poincaré algebra
\cite{Harmark:2019upf}.%
\footnote{The geometrization of higher $p$-forms was pioneered in \cite{DAuria:1982uck} in the context of supergravities. See also
\cite{DAuria:2020guc} for a recent review.
It would be interesting to explore the relation between our construction both to this formalism as well as to the vielbeine of double field theory~\cite{Hohm:2010pp}.}
In the same spirit as doubled field theory \cite{Hull:2009mi,Hohm:2010pp}, this includes a second set of translational generators that give rise to an extra gauge connection which enters in the parametrization of the Kalb--Ramond field. From an algebraic point of view, our FSG algebra arises as an İnönü--Wigner contraction of this string Poincaré algebra.

Since the NR string action is manifestly invariant under all the gauge symmetries, the background geometry does
not need to be supplemented by a torsion/foliation constraint.
This is to be contrasted to the situation~\cite{Bergshoeff:2018yvt} where there is a $Z_A$ symmetry leading to a torsion constraint on the longitudinal vielbeine.
As emphasized above, the TSNC geometry arises naturally once we take into account that fundamental strings
are extremal under the Kalb--Ramond field, and its derivation is entirely analogous to how TNC geometry arises
from the large speed of light limit of extremal particles. We have furthermore verified that the geometry and its symmetries can also be nicely reproduced by a null reduction of fundamental strings.

The results of this paper allow for a variety of further interesting lines of investigation.
Firstly, two important further  directions to address are beta functions and effective spacetime actions in our TSNC formulation. In this paper, we have shown that the torsion of the target space can be seen as decoupled from the one-form gauge symmetry of $m_{\mu\nu}$, which is natural as $m_{\mu\nu}$ does not couple to $\tau^A_\mu$ in our Nambu-Goto action. However, in the  beta-function calculations of
\cite{Gomis:2019zyu,Gallegos:2019icg,Bergshoeff:2019pij,Yan:2019xsf,Gallegos:2020egk,Bergshoeff:2021bmc}  one finds that certain restrictions on the target space geometry are needed in order keep the non-relativistic target space geometry from flowing back to the relativistic geometry. Since our analysis is classical, it is not sensitive to this. Additionally, one should impose a general torsion constraint to ensure that it is possible to foliate the geometry. It would be highly interesting to investigate how these constraints on the geometry are connected to each other. In addition to this,
spacetime actions from non-relativistic limits were considered for the NS-NS string in the recent paper \cite{Bergshoeff:2021bmc}. It would be interesting to reexamine this using TSNC geometry.

Another related direction to investigate is whether one can include the dilaton field of the NS-NS sector from a suitable extension of string Poincaré geometry and subsequently apply the non-relativistic limit.

A different direction, which was among our original motivations for this work, is to understand the general target space for the non-relativistic sigma-models that one finds via the Spin Matrix theory limits \cite{Harmark:2014mpa}. In \cite{Harmark:2017rpg,Harmark:2018cdl}
(see also \cite{Harmark:2019upf,Harmark:2020vll}), the target space was identified
as a $U(1)$-Galilean geometry for a special case. With our new understanding of the target space geometry of the NR string, we can identify the target space geometry that follows when one takes the non-relativistic worldsheet  limit that corresponds to the Spin Matrix limit. This will be investigated in our upcoming paper \cite{GCA} where we shall also perform a Hamiltonian analysis of the resulting worldsheet theory.

It would obviously also be interesting to generalize the method developed in this paper to the non-relativistic limit of extremal branes.
This will be addressed in our follow-up paper \cite{NRB},
in which we will present the analogue of TSNC for $p$-branes and the underlying symmetry algebra.
Since D/M-branes  are such objects, albeit with additional worldvolume fields and a specific dilaton coupling, this will be relevant for the geometry of these objects in non-relativistic string/M-theory.%
\footnote{See e.g. \cite{Kluson:2021pux,Kluson:2021djs}
and \cite{Blair:2021ycc}
for recent work on non-relativistic M-branes using
null reduction and limits respectively.}
In addition, it would be useful to connect to the results on the non-relativistic open string sector and DBI actions obtained in \cite{Gomis:2020fui,Gomis:2020izd}.

Finally, another natural direction to pursue is the supersymmetric generalization of string Poincaré symmetry,
providing insights into how R-R fields can be geometrically captured. The latter will likely be related to exceptional geometry
just as TSNC geometry is connected to doubled geometry.
This will be interesting in its own right, while the NR limit of such a geometry will likely play a role in D/M-brane Newton--Cartan geometry and also show a more complete picture of NR R-R fields and non-perturbative dualities in NR string theory.

\section*{Acknowledgements}

We thank Eric Bergshoeff, Emil Have, Jan Rosseel and Ziqi Yan for useful discussions. The work of JH is supported by the Royal Society University Research Fellowship ``Non-Lorentzian Geometry in Holography'' (grant number UF160197). The work of LB is supported by the Royal Society Research Fellows Enhancement Award 2017 “Non-Relativistic Holographic Dualities” (grant number RGF$\backslash$EA$\backslash$180149). The work of TH, NO and GO is supported in part by the project ``Towards a deeper understanding of  black holes with non-relativistic holography'' of the Independent Research Fund Denmark (grant number DFF-6108-00340) and the work of NO and GO also by the Villum Foundation Experiment project 00023086.

\appendix

\section{Review of the non-relativistic particle}
\label{sec:particle}

In this appendix we review how one can obtain the coupling of a non-relativistic particle to torsional Newton--Cartan (TNC) geometry from a null reduction or an extremal limit.
This serves as background and intuition for Sections \ref{sec:null_reduction} and \ref{sec:string_limit}, respectively,

\subsection{Null reduction}
\label{sec:particle_nullred}
One way to obtain the coupling of a non-relativistic particle to a TNC background in $D$ spacetime dimensions is through null reduction of a massless relativistic particle in $D+1$ dimensions.
Parametrizing the null isometry of the relativistic $(D+1)$-dimensional geometry with $\pd_u$, we can write the metric as
\begin{equation}
  \label{nullreduction}
  ds^2
  = g_{MN} dX^M dX^N
  = 2\tau_\mu (du - m_\mu dx^\mu) + h_{\mu \nu} dx^\mu dx^\nu  \,,
\end{equation}
where none of the metric components depend on $u$.
Here, $\mu = 0,\ldots,D-1$ are the lower-dimensional indices.
Furthermore, we can introduce the spatial vielbein decomposition $h_{\mu \nu} = e_\mu^a e_\nu^b \delta_{ab}$ with $a=1,\ldots,D-1$ as spatial frame indices.
On such a background, the relativistic massless particle action is
\begin{equation}
  S
  = \int \frac{1}{2e}g_{MN}\dot X^M\dot X^N d \lambda
  = \int \left(
    \frac{1}{e}\dot{u}\tau_\mu \dot X^\mu
    + \frac{1}{2e}\bar h_{\mu\nu}\dot X^\mu \dot X^\nu
  \right) d\lambda\,,
\end{equation}
where we have defined $\bar{h}_{\mu\nu} = h_{\mu\nu} - \tau_\mu m_\nu - \tau_\nu m_\mu$ and where the dot denotes differentiation with respect to the worldline parameter $\lambda$.
The momentum $p_u = \pd \LL/ \pd \dot{u} = \tau_\mu \dot{X}^\mu / e$ is conserved, so we can set $p_u = m$ and solve this for $e = \tau_{\mu}\dot{X}^\mu/ p_u$.
The action then is
\begin{equation}
  \label{eq:app-particleaction}
  S
  = \frac{m}{2} \int \frac{\bar{h}_{\mu\nu} \dot{X}^\mu \dot{X}^\nu}{\tau_\rho \dot{X}^\rho} d\lambda\,,
\end{equation}
which is the action for a non-relativistic point particle of mass $m$.

The decomposition of the line element in \eqref{nullreduction} admits the following local symmetries
\begin{subequations}
  \label{particletrafos}
  \begin{gather}
    \bar \delta \tau_\mu =  \CL_\xi \tau_\mu
    \spa
    \delta e^a_\mu = \CL_\xi e^a_\mu+   \lambda^a \tau_\mu + \lambda^a{}_b e^b_\mu
    \spa
    \delta h_{\mu \nu} = \CL_\xi  h_{\mu \nu}
    + 2\lambda_{a}  \tau_{(\mu } e^a_{ \nu)} \,,
    \\
    \bar \delta m_{\mu } = \CL_\xi m_{\mu } + \lambda_a e^a_{\mu} + \partial_\mu \sigma\,,
  \end{gather}
\end{subequations}
corresponding to diffeomorphisms $\xi^\mu$, local rotations $\lambda_{ab}=-\lambda_{ba}$, local Galilean boosts $\lambda_a$ and the Bargmann $U(1)$ transformations parametrized by $\sigma$.
Here, spatial frame indices are raised and lowered with $\delta^{ab}$ and $\delta_{ab}$.
It is easy to check that the action \eqref{eq:app-particleaction} is invariant under these transformations.
The background geometry that the action couples to is
\begin{equation}
  \mbox{torsional Newton--Cartan (TNC) geometry:} \quad  \tau_\mu \spa h_{\mu \nu} \spa m_\mu\,,
\end{equation}
whose transformations are given by~\eqref{particletrafos}.
For TNC geometry, the intrinsic torsion of metric-compatible connections is parametrized~\cite{Christensen:2013lma,Figueroa-OFarrill:2020gpr} by $d\tau$.
The case of zero torsion $d\tau =0$ corresponds to absolute time and this is the geometry that allows us to formulate Newtonian gravity in a frame-independent way, at least on shell,\footnote{An off-shell formulation was found in \cite{Hansen:2020pqs} from the large speed of light expansion of GR, providing an action formulation of Newtonian gravity.}
as pioneered by Cartan.
On the other hand, in twistless torsional Newton--Cartan (TTNC) geometry, we have $\tau \wedge d \tau =0$ which is equivalent to the condition of hypersurface orthogonality, so that the spacetime admits a foliation in terms of equal-time slices.
The derivation above shows that, in the null reduction perspective on the TNC particle action, there is no constraint on the torsion.

For strings, the null reduction procedure was applied in~\cite{Harmark:2017rpg} to the Nambu-Goto action, and it was later generalized to a Polyakov action, including the $B$-field coupling, in \cite{Harmark:2018cdl,Harmark:2019upf}.
The result is the action \eqref{null_SNC_action} describing non-relativistic strings coupling to TNC geometry.
Just as for the particle case reviewed above, no torsion constraint appears in this construction.
We revisit this action in Section~\ref{sec:null_reduction} and provide a new perspective on its symmetries.

\subsection{Newton--Cartan geometry from gauging the Bargmann algebra}
\label{sec:particle_gauging}
As we mentioned in Section~\ref{sec:string_limit}, one can obtain the transformation of a pseudo-Riemannian metric from a gauging procedure of the Poincaré group.
In this construction, the vielbein corresponds to the gauge field of translations and the spin connection is gauge field of Lorentz transformations.
Gauging the Lorentz transformations ensures local Lorentz symmetry, as required by the Einstein equivalence principle.
To obtain diffeomorphisms from local translations without constraining the torsion of the geometry, one can use the $\bar\delta$ transformations defined in Section~\ref{sec:gaugingKR}.

Likewise, non-relativistic geometry can be defined from an algebraic point of view by gauging a non-relativistic spacetime algebra.
In particular, Newton--Cartan geometry and its transformations~\eqref{particletrafos} can be obtained from the Bargmann algebra~\cite{Andringa:2010it}, which has generators $H$ (Hamiltonian) $P_a$ (translations), $J_{ab}$ (rotations), $G_a$ (Galilean boosts) and $N$ (mass operator)
and non-zero commutation relations
\begin{subequations}
  \label{eq:app-bargmann}
  \begin{align}
    [J_{ab},P_c] & =  2 \delta_{c[a} P_{b]}  \spa
    [J_{ab}, G_{c}] = 2 \delta_{c[a} G_{b]}     \spa
    [J_{ab}, J_{cd}]
    = 2\delta_{c[a} J_{b]d} - (c \leftrightarrow d)\,,  \\
    [G_{a}, P_b]
    &=  - \delta_{ab} N  \spa
    [G_{a}, H] = - P_a\,.
  \end{align}
\end{subequations}
The details of the gauging procedure can be found in Appendix~\ref{app:fsnc-gauging}, where we apply it to a more involved algebra.
From a null reduction perspective, the Bargmann algebra appears as the centralizer of the null translation generator $P_+$ in the higher-dimensional Poincaré algebra.

\subsection{Non-relativistic particle from the limit of an extremal particle}
\label{sec:particle_limit}
Alternatively, the action~\eqref{eq:app-particleaction} can be obtained from a non-relativistic limit of a charged relativistic point particle, see for example~\cite{Jensen:2014wha,Bergshoeff:2015uaa}.
In contrast to the null reduction procedure, the resulting non-relativistic particle couples to the same number of dimensions as the relativistic particle we start with.
One reason to start with a charged particle is that, for a given spacetime dimension, the Bargmann algebra~\eqref{eq:app-bargmann} has one more generator than the Poincaré algebra, which can be constructed using an extra $U(1)$ gauge field.
As we will see, this field will also allow us to absorb divergences that would otherwise arise in the $c\to\infty$ limit.

In this case, our starting point is the action for a charged relativistic particle,
\begin{equation}
  S =  - mc \int \sqrt{- g_{\mu\nu} \dot{x}^\mu \dot{x}^\nu} d\lambda
  + q \int A_\mu \dot{x}^\mu d\lambda\,.
\end{equation}
The first step is to decompose the metric in timelike and spacelike components,
\begin{equation}
  g_{\mu \nu} = -c^2 T_\mu T_\nu + h_{\mu\nu}\,.
\end{equation}
The explicit factor of $c^2$ means that the light cone opens up in the $c\rightarrow \infty$ limit.
As before, we can introduce spatial vielbeine $e_\mu^a$ corresponding to $h_{\mu \nu} =  e_\mu^a e_\nu^b \delta_{ab}$.
Expanding the action for large $c$, assuming $T_\mu\dot x^\mu>0$, we find
\begin{equation}
  \label{eq:app-pp-expanded-iw-action}
  S
  = - mc^2 \int \left[  T_\mu -  \frac{q}{mc^2} A_\mu \right] \dot x^\mu d\lambda
  + \frac{m}{2} \int \frac{ h_{\mu \nu}  \dot x^\mu  \dot x^\nu }{T_\rho  \dot x^\rho} d\lambda
  + \CO (c^{-2})\,.
\end{equation}
Let us now consider an extremal particle with $q=mc^2$.
We then make the following $c$-dependent basis transformation before taking the limit
\begin{subequations}
  \label{TAdef}
  \begin{align}
    T_\mu
    &= \tau_\mu + \frac{1}{2c^2} m_\mu\,,
    \\
    A_\mu
    &= \tau_\mu   -  \frac{1}{2c^2} m_\mu\,,
  \end{align}
\end{subequations}
Substituting these redefinitions in the expanded action~\eqref{eq:app-pp-expanded-iw-action} for an extremal particle, we see that the leading term vanishes.
In the $c\rightarrow \infty$ limit, we recover precisely the action~\eqref{eq:app-particleaction} of a neutral non-relativistic particle probing TNC geometry.
As in the null reduction construction of Section~\ref{sec:particle_nullred}, there is no constraint on the torsion $d\tau$ of the background TNC geometry.

For completeness we also discuss the symmetry algebra from the limit point of view.
For the charged relativistic particle this is Poincaré plus a $U(1)$ algebra, whose generator we call $Q$.
We can implement the basis transformation \eqref{TAdef} on the algebra, giving
\begin{equation}
\label{HNcombo}
  H = c P_0 + Q
  \spa
  N = \frac{1}{2c^2} \left( c P_0 - Q\right)\,.
\end{equation}
Taking the $c\rightarrow \infty$ limit (corresponding to an İnönü--Wigner contraction), the resulting algebra is the Bargmann algebra~\eqref{eq:app-bargmann}, where in particular $H$ and $N$ are the generators corresponding to the gauge fields $\tau_\mu$ and $m_\mu$ introduced in Equation~\eqref{TAdef} above.

We thus conclude that the limiting procedure applied to an extremal relativistic particle provides an alternative route to get the action of a neutral non-relativistic particle to TNC geometry.
In the main text of this paper, we  mimic this exact procedure to obtain the action for non-relativistic strings.
To identify the analogue of TNC geometry, starting from fundamental strings, one crucially needs to consider how the $U(1)$ factor above generalizes for extended objects.

As a final remark, one could wonder what the action of a {\it charged} non-relativistic particle coupling to TNC geometry is.
For this, one can simply add a coupling $q_{NR} \int A$ to the action~\eqref{eq:app-particleaction}.
As is well known, since the total action then depends only on the combination $m\, m_\mu + q A_\mu$, there is an extra shift (or Stückelberg) symmetry that leaves the action invariant.
The same action can also be obtained by keeping $T_\mu$ and $A_\mu$ general, i.e.
by not performing the transformation \eqref{TAdef} and taking the point of view of expanding the action as opposed to a limit with a vanishing divergent term (see also \cite{Hansen:2020pqs}).

\section{Details of non-relativistic string algebras}
\label{app:algebras}
In this appendix, we provide further details on the non-relativistic algebras relevant to the main text.
First, in Section~\ref{app:snc-fsnc-algebras}, we review the string Galilei algebra, which is obtained from a $c\to\infty$ limit of the Poincaré algebra where not one but two directions are distinguished.
The string Galilei algebra can be extended in several ways, and perhaps the most prominent one leads to what is known as the string Newton--Cartan (SNC) algebra.
This algebra is closely related to the non-relativistic string action~\eqref{SNC_action} that we discussed in the Introduction.
In addition, we consider the F-string Galilei (FSG) algebra, which extends the string Galilei algebra by adding generators corresponding to local one-form transformations, and we compare the two in Section~\eqref{eq:app-mdef-repeat}

Finally, in Section~\ref{app:fsnc-gauging}, we discuss the gauging of the FSG algebra.
We present a procedure that reproduces both diffeomorphisms and local one-form gauge transformations.
As discussed in Section~\ref{sec:symmetries_TSNC}, these symmetries are realized in our non-relativistic string action~\eqref{NRaction2} without imposing any constraints on the torsion of the target space TSNC geometry.
This is in contrast to the usual realization~\cite{Bergshoeff:2018vfn,Bergshoeff:2019pij} of the SNC algebra in the non-relativistic string action~\eqref{SNC_action}, which requires zero torsion.
In the quantum theory~\cite{Gallegos:2019icg,Bergshoeff:2019pij,Gomis:2019zyu,Yan:2019xsf}, such torsion constraints play an important role, but zero torsion may be overly restrictive~\cite{Harmark:2019upf,Bergshoeff:2021bmc}.
Instead of being imposed at the level of the action, constraints on the torsion may also be generated dynamically as part of the equations of motion, and these two perspectives were recently explored in~\cite{Yan:2021lbe}.

\subsection{String Galilei, string Newton--Cartan and F-string Galilei algebras}
\label{app:snc-fsnc-algebras}
Our starting point is the Poincaré algebra,
\begin{subequations}
  \label{eq:app-poincare-commutation-relations}
  \begin{align}
    [M_{\ab\bb}, M_{\cb,\db}]
    &= \eta_{\ab\cb} M_{\bb\db}
    - \eta_{\bb\cb} M_{\ab\db}
    + \eta_{\bb\db} M_{\ab\cb}
    - \eta_{\ab\db} M_{\bb\cb}\,,
    \\
    [M_{\ab\bb}, P_\cb]
    &= \eta_{\ab\cb} P_\bb - \eta_{\bb\cb} P_\ab\,.
  \end{align}
\end{subequations}
Here, the indices $\ab,\bb,\ldots$ label $D$ spacetime dimensions.
The generators $M_{\ab\bb}$ and $P_\ab$ correspond to Lorentz transformations and translations.

We can obtain the string Galilei algebra by decomposing $\ab=(A,a)$, which singles out one space and one time direction $A=0,1$, as well as $D-2$ spatial directions $a=2,\ldots,D-1$.
These are referred to as the longitudinal and transverse directions, respectively.
Then defining the generators
\begin{equation}
  H_A = c P_A\,,
  \quad
  P_a\,,
  \quad
  Q_a\,,
  \quad
  J_{AB} = \epsilon_{AB}J = M_{AB}\,,
  \quad
  J_{ab} = M_{ab}\,,
  \quad
  G_{Ab} = \frac{1}{c} M_{Ab}\,,
\end{equation}
and taking the $c\to\infty$ limit, we obtain the \emph{string Galilei algebra}~\cite{Brugues:2004an,Andringa:2012uz},
\begin{subequations}
\label{eq:app-string-galilei-algebra}
  \begin{align}
    [J_{ab}, J_{cd}]
    &= \delta_{ac} J_{bd}
    - \delta_{bc} J_{ad}
    + \delta_{bd} J_{ac}
    - \delta_{ad} J_{bc}\,,
    \\
    [J,G_{Ab}]
    &= \epsilon^C{}_A G_{Cb}\,,
    \\
    [J_{ab}, G_{Cd}]
    &= \delta_{ad} G_{Cb} - \delta_{bd} G_{Ca}\,,
    \\
    [J, H_A]
    &= \epsilon^B{}_A H_B\,,
    \\
    [G_{Ab}, H_C]
    &= \eta_{AC} P_b\,,
    \\
    [J_{ab},P_c]
    &= \delta_{ac} P_b - \delta_{bc} P_a\,.
  \end{align}
\end{subequations}
For the non-relativistic particle, we need to consider the (central) extension of the Galilei algebra given by the Bargmann algebra~\eqref{eq:app-bargmann} in order to obtain a massive representation.
For the string actions considered in this paper, we likewise have to consider (non-central) extensions of the string Galilei algebra.

One notable candidate is known as the \emph{string Newton--Cartan (SNC) algebra}~\cite{Brugues:2004an,Andringa:2012uz}.
It introduces the extensions $Z_A$ and $Z_{AB}$, and its commutation relations are given by~\eqref{eq:app-string-galilei-algebra} together with the following additional commutation relations
\begin{subequations}
  \label{eq:app-snc-algebra-additional-commutators}
  \begin{align}
    [G_{Ab},G_{Cd}]
    &= \delta_{bd} Z_{AC}\,,
    \\
    [J, Z_A]
    &= \epsilon^B{}_A Z_B\,,
    \\
    [G_{Ab}, P_c]
    &= - \delta_{bc} Z_A\,,
    \\
    [Z_{AB}, H_C]
    &= \eta_{AC} Z_B - \eta_{BC} Z_A\,.
  \end{align}
\end{subequations}
Here, we take $Z_{AB}$ to be antisymmetric.\footnote{%
  For simplicity, we take the $Z_{AB}$ generator to be antisymmetric.
  In for example~\cite{Bergshoeff:2019pij}, the SNC algebra instead refers to a generalization where $Z_{AB}$ is traceless but not necessarily antisymmetric $Z_{AB}$, and the case with antisymmetric $Z_{AB}$ is referred to as the string Bargmann algebra.
}
This algebra can be obtained as a quotient of an expansion of the Poincaré algebra~\cite{Harmark:2019upf}, see also~\cite{Bergshoeff:2019ctr} for a related construction in four spacetime dimensions.
It would be interesting to understand if the SNC algebra can be obtained from a contraction.
In their standard realization, the transformations associated to $Z_A$ are only symmetries of the associated non-relativistic string action~\eqref{SNC_action} if the constraint $D_{[\mu} \tau^A_{\nu]} = 0$ is imposed, where $\tau^A$ is the longitudinal vielbein associated to $H_A$.
The symmetries of the corresponding non-relativistic string action~\eqref{SNC_action} consist of the transformations that are obtained by gauging the SNC algebra, and in addition contain the one-form gauge transformations of the two-form gauge potential, which is not described by the SNC algebra.

As discussed in Section~\ref{sec:string_limit}, we can incorporate the one-form gauge transformations of the Kalb--Ramond field in the relativistic theory from the string Poincaré algebra~\eqref{eq:poincare-plus-vector-commutation-relations},
\begin{subequations}
  \label{eq:app-poincare-plus-vector-commutation-relations-repeat}
  \begin{align}
    [M_{\ab\bb}, M_{\cb\db}]
    &= \eta_{\ab\cb} M_{\bb\db}
    - \eta_{\bb\cb} M_{\ab\db}
    + \eta_{\bb\db} M_{\ab\cb}
    - \eta_{\ab\db} M_{\bb\cb}\,,
    \\
    [M_{\ab\bb}, P_\cb]
    &= \eta_{\ab\cb} P_\bb - \eta_{\bb\cb} P_\ab\,,
    \\
    [M_{\ab\bb}, Q_\cb]
    &= \eta_{\ab\cb} Q_\bb - \eta_{\bb\cb} Q_\ab\,.
  \end{align}
\end{subequations}
In Section~\ref{sec:gaugingKR}, building on~\cite{Harmark:2019upf}, we show how this algebra can be gauged to construct NS-NS geometry consisting of Lorentzian vielbeine, a spin connection and a Kalb--Ramond field, together with corresponding diffeomorphism symmetry, local Lorentz transformations and one-form gauge transformations.
(We do not consider the dilaton in this work.)
Starting from this algebra, we can then define the generators
\begin{subequations}
  \begin{gather}
    \label{eq:app-HNcombo-strings}
    H_A = c (P_A +  Q_B \epsilon^B{}_A)\,,
    \quad
    N_A = \frac{1}{2c} (\epsilon_A{}^B  P_B +  Q_A)\,,
    \\
    P_a\,,
    \quad
    Q_a\,,
    \quad
    J_{AB} = \epsilon_{AB}J = M_{AB}\,,
    \quad
    J_{ab} = M_{ab}\,,
    \quad
    G_{Ab} = \frac{1}{c} M_{Ab}\,.
  \end{gather}
\end{subequations}
Note the similarity of the redefinitions~\eqref{eq:app-HNcombo-strings} to the generators~\eqref{HNcombo} in the definition of the Bargmann algebra above.
In the $c\to\infty$ limit, this leads to the commutation relations~\eqref{eq:app-string-galilei-algebra} of the string Galilei algebra, supplemented with/replaced by
\begin{subequations}
  \label{eq:app-tsnc-algebra-additional-commutators}
  \begin{align}
    [J, N_A]
    &= \epsilon^B{}_A N_B\,,
    \\
    [G_{Ab}, H_C]
    &= \eta_{AC} P_b + \epsilon_{AC} Q_b\,,
    \\
    [G_{Ab}, P_c]
    &= \delta_{bc} \epsilon^B{}_A N_B\,,
    \\
    \label{eq:app-fsnc-q-a-not-ideal}
    [G_{Ab}, Q_c]
    &= - \delta_{bc} N_A,
    \\
    [J_{ab},Q_c]
    &= \delta_{ac} Q_b - \delta_{bc} Q_a\,.
  \end{align}
\end{subequations}
which we refer to as the \emph{F-string Galilei (FSG) algebra}.
As we will show in Section~\eqref{app:fsnc-gauging}, the corresponding geometry that results from gauging this algebra can have arbitrary torsion, just as torsional Newton--Cartan (TNC) geometry arises from the gauging of the Bargmann algebra~\eqref{eq:app-bargmann}.

\subsection{Comparing SNC and TSNC geometries}
\label{app:comparing-snc-tsnc}
At first sight, the additional commutators~\eqref{eq:app-tsnc-algebra-additional-commutators} that define the FSG algebra with respect to the string Galilei algebra~\eqref{eq:app-string-galilei-algebra} are similar to the additional commutators~\eqref{eq:app-snc-algebra-additional-commutators} that define the SNC algebra.

First, the SNC algebra contains the central extension $Z_{AB}$, whereas the FSG algebra contains the non-central generators $Q_c$ that are related to one-form gauge transformations.
Both generators are crucial to the Jacobi identities of their respective algebras, and it is easy to see from~\eqref{eq:app-snc-algebra-additional-commutators} and~\eqref{eq:app-tsnc-algebra-additional-commutators} that they do not form an ideal and hence cannot be quotiented out consistently.
Also, in SNC geometry, the Kalb--Ramond field and the associated one-form gauge transformations are not related to the algebra, while both follow from the gauging of the FSG algebra, as we will discuss in detail in Section~\ref{app:fsnc-gauging}.

On the other hand, the SNC generator $Z_A$ is similar to $\epsilon_A{}^B N_B$ in the FSG algebra.
From the expression for the inverse vielbein $\tau^\alpha_A$ in Equation~\eqref{dettau}, we can derive the following identity for an arbitrary matrix $C^A_\alpha$,
\begin{subequations}
  \label{eq:stueckelberg-identity}
  \begin{align}
    \epsilon^{\alpha\beta} \eta_{AB} \tau^A_{[\alpha} C^B_{\beta]}
    &= - \sqrt{-\tau} \epsilon^A{}_B \tau^\alpha_A C^B_\alpha
    \\
    &= - \sqrt{-\tau} \eta^{AB} \tau_A^\alpha \epsilon_{BC} C^C_\alpha
    \\
    &= - \sqrt{-\tau} \eta^{AB} \tau_A^\alpha \tau_B^\beta
    \epsilon_{CD} \tau_{(\alpha}^C C^D_{\beta)}.
  \end{align}
\end{subequations}
In the main text, we used this identity for $C^A_\alpha=\pi^A_\alpha$ together with
$\pi^A_\alpha = - 2\epsilon^A{}_B m^B_\alpha$
to rewrite the non-relativistic string action~\eqref{NRaction2}
\begin{equation}
  \label{eq:app-NRaction2-repeat}
  S_{\rm NR}
  =  -\frac{T}{2} \int d^2 \sigma \left[
    \sqrt{-\tau} \, \eta^{AB} \tau_A^\alpha \tau_B^\beta h_{\alpha\beta} +
    \epsilon^{\alpha\beta} m_{\alpha\beta}
  \right]\,,
\end{equation}
in the alternative form~\eqref{NRactionSNC},
\begin{equation}
  \label{eq:app-NRactionSNC-repeat}
  S_{\rm NR} =  -\frac{T}{2} \int d^2 \sigma \left[ \sqrt{-\tau} \, \eta^{AB} \tau_A^\alpha \tau_B^\beta \bar h_{\alpha\beta} +
  \epsilon^{\alpha\beta} B^\nparallel_{\alpha\beta} \right]\,,
\end{equation}
where we recall that $B^\nparallel_{\mu \nu}=\delta_{ab} e^a_{[\mu} \pi^b_{\nu]}$ denotes the non-longitudinal part of $m_{\mu\nu}$ and
\begin{gather}
  m_{\mu \nu}
  = \eta_{AB} \tau^A_{[\mu} \pi^B_{\nu]} + \delta_{ab} e^a_{[\mu} \pi^b_{\nu]}\,,
  \\
  \bar h_{\mu \nu}
  =  h_{\mu \nu} - \frac{1}{2} \epsilon_{AB} ( \tau_\mu^A \pi_\nu^B +  \tau_\nu^A \pi_\mu^B )
  =  h_{\mu \nu} + \eta_{AB} \left( \tau_\mu^A m_\nu^B +  \tau_\nu^A m_\mu^B \right)
  \,,
\end{gather}
However, in the literature on SNC geometry~\cite{Bergshoeff:2018yvt,Bergshoeff:2019pij}, the string action is often taken to be as in Equation~\eqref{SNC_action},
\begin{equation}
  \label{eq:app-SNC_action-repeat}
  S=- \frac{T}{2} \int d^2\sigma
  \Big[ \sqrt{-\tau} \, \tau^{\alpha\beta} \bar{h}_{\mu\nu} + \epsilon^{\alpha\beta} B_{\mu\nu} \Big]
  \partial_\alpha X^\mu \partial_\beta X^\nu\,,
\end{equation}
where $B_{\mu\nu}$ is fully general and not necessarily non-longitudinal.
Crucially, without that restriction on $B_{\mu\nu}$, this means that the action~\eqref{eq:app-SNC_action-repeat} has a Stückelberg symmetry~\cite{Bergshoeff:2018yvt,Harmark:2019upf,Bergshoeff:2019pij} corresponding to the $C^A_\alpha$ in Equation~\eqref{eq:stueckelberg-identity}
that transforms part of the $B_{\mu\nu}$ coupling into the $\bar{h}_{\mu\nu}$ coupling.
We can fix this redundancy is by imposing $B_{\mu\nu}=B^\nparallel_{\mu\nu}$, which recovers the action~\eqref{eq:app-NRactionSNC-repeat}.
Alternatively, we can keep $B_{\mu\nu}$ general but fix $m^A_\mu=0$, which reproduces the action~\eqref{eq:app-NRaction2-repeat} upon identifying $B_{\alpha\beta}=m_{\alpha\beta}$.

Furthermore, note that the Galilean boost generators $G_{Ab}$ commute in the FSG algebra but not in the SNC algebra.
As we can see from the following section, and as we mentioned in Equations~\eqref{eq:gauging-fsnc-boost-deltabar-transformations} and~\eqref{eq:tsnc-boost-composite-fields}, the boosts act on the TSNC variables as follows,
\begin{subequations}
  \begin{align}
    \bar\delta \tau^A_\mu
    &= 0,
    \\
    \bar\delta e^a_\mu
    &= - \lambda_B{}^a \tau^B_\mu\,,
    \\
    \bar\delta m_{\mu\nu}
    &= 2\epsilon_{AB} \lambda^A{}_c \tau^B_{[\mu} e^c_{\nu]}\,.
  \end{align}
\end{subequations}
As a result, two successive boosts on $e^a_\mu$ give zero, and on $m_{\mu\nu}$ they give
\begin{equation}
  \delta_2 \delta_1 m_{\mu\nu}
  = 2\epsilon_{AB} \lambda_1^A{}_c \tau^B_{[\mu} \delta_2 e^c_{\nu]}
  = - 2 \epsilon_{AB} \lambda_1^A{}_c \lambda_{2 D}{}^c \tau^B_{[\mu} \tau^D_{\nu]}\,.
\end{equation}
The commutation of two boosts then gives
\begin{equation}
  \label{eq:fsnc-two-boosts-on-m-mu-nu}
  \left( \delta_2 \delta_1 - \delta_1 \delta_2 \right) m_{\mu\nu}
  = 2 \epsilon_{AB} \left(
    \lambda_1^B{}_c \lambda_{2 D}{}^c
    - \lambda_2^B{}_c \lambda_{1 D}{}^c
  \right)
  \tau^A_{[\mu} \tau^D_{\nu]}
  = 0\,,
\end{equation}
which reflects the fact that $[G_{Ab},G_{Cd}]=0$ in the FSG algebra.
On the other hand, one can show that in SNC geometry, we have
\begin{subequations}
  \begin{align}
    \bar\delta \tau^A_\mu
    &= 0,
    \\
    \bar\delta e^a_\mu
    &= - \lambda_B{}^a \tau^B_\mu,
    \\
    \bar\delta m^A_\mu
    &= \lambda^A{}_b e^b_\mu,
  \end{align}
\end{subequations}
which results in the following non-zero commutator of two boosts acting on $m^A_\mu$,
\begin{equation}
  \left(\delta_2 \delta_1 - \delta_1 \delta_2\right) m^A_\mu
  = \left(\lambda_1^A{}_b \lambda_{2 C}{}^b - \lambda_2^A{}_b \lambda_{1 C}{}^b\right) \tau^B_\mu.
\end{equation}
This corresponds to the non-zero commutator $[G_{Ab}, G_{Cd}]= \delta_{bd} Z_{AC}$ in the SNC algebra~\eqref{eq:app-snc-algebra-additional-commutators}.
However, note that this is only manifested on the $m_\mu^A$ field, so if we use the Stückelberg symmetry~\eqref{eq:stueckelberg-identity} to set $m^A_\mu=0$, we see that $Z_{AB}$ is effectively removed from the algebra.
As a result, we can conclude that with the Stückelberg gauge fixing as outlined above, the SNC string action~\eqref{eq:app-SNC_action-repeat} reduces to the TSNC action~\eqref{eq:app-NRaction2-repeat}.

Finally, as we mentioned at the start of this appendix, the standard realization of the SNC $Z_A$ transformations in the string action~\eqref{eq:app-SNC_action-repeat} requires that $D_{[\mu} \tau^A_{\nu]}=0$, which can be interpreted as a constraint on the torsion of the SNC geometry.
As we will see in Section~\ref{eq:app-string-galilei-algebra} below, the $N_A$ and $Q_a$ transformations in the FSG algebra reproduce the one-form gauge transformations of $m_{\mu\nu}$, which is a symmetry of the action~\eqref{eq:app-NRaction2-repeat} without imposing any constraints on $\tau^A$ or the torsion.
For this reason, the corresponding geometry allows for arbitrary torsion, which is the reason why we refer to the geometry and the algebra as torsional Newton--Cartan.

\subsection{Gauging the F-string Galilei algebra}
\label{app:fsnc-gauging}
Now let us consider the gauging of the FSG algebra defined by the commutators~\eqref{eq:app-string-galilei-algebra} and~\eqref{eq:app-tsnc-algebra-additional-commutators}, see also Equation~\eqref{eq:fsnc-algebra-dual-rotations} in the main text.
We introduce the FSG-valued connection and the transformation parameter
\begin{align}
  \mA_\mu
  &= \tau^A_\mu H_A + e^a_\mu P_a
  + \omega_\mu J + \frac{1}{2} \omega_\mu{}^{ab} J_{ab} + \omega_\mu{}^{Ab} G_{Ab}
  + \pi^A_\mu N_A + \pi^a_\mu Q_a\,,
  \\
  \Lambda
  &= \zeta^A H_A + \zeta^a P_a
  + \sigma J + \frac{1}{2} \sigma^{ab} J_{ab} + \sigma^{Ab} G_{Ab}
  + \kappa^A N_A + \kappa^a Q_a\,.
\end{align}
Here, $\tau^A_\mu$ and $e^a_\mu$ are longitudinal and transverse vielbeine,
while $\omega_\mu$, $\omega_\mu{}^{ab}$ and $\omega_\mu{}^{Ab}$ are spin connections
and $\pi^A_\mu$ and $\pi^a_\mu$ are additional gauge fields associated to the $m_{\mu\nu}$ field introduced in Equation~\eqref{eq:mdef}, which is defined by
\begin{equation}
  \label{eq:app-mdef-repeat}
  m_{\mu\nu}
  = \eta_{AB} \tau^A_{[\mu} \pi^B_{\nu]}
  + \delta_{ab} e^a_{[\mu} \pi^b_{\nu]}\,.
\end{equation}
The transformations of these fields are determined by the adjoint gauge transformation $\delta \mA_\mu = \pd_\mu \Lambda + [\mA,\Lambda]_\mu$, which gives
\begin{subequations}
  \label{eq:tsnc-adjoint-transformations}
  \begin{align}
    \label{eq:tsnc-adjoint-transformations-tau-A}
    \delta \tau^A_\mu
    &= \pd_\mu \zeta^A + \omega_\mu \epsilon^A{}_B \zeta^B
    - \sigma \epsilon^A{}_B \tau^B_\mu\,,
    \\
    \delta e^a_\mu
    &= \pd_\mu \zeta^a - \omega_\mu{}^a{}_b \zeta^b + \omega_{\mu B}{}^a \zeta^B
    + \sigma^a{}_b e^b_\mu - \sigma_B{}^a \tau^B_\mu\,,
    \\
    \delta \omega_\mu
    &= \pd_\mu \sigma\,,
    \\
    \delta \omega_\mu{}^{ab}
    &= \pd_\mu \sigma^{ab} - \omega_\mu{}^a{}_c \sigma^{cb}
    - \omega_\mu{}^b{}_c \sigma^{ac}\,,
    \\
    \delta \omega_\mu{}^{Ab}
    &= \pd_\mu \sigma^{Ab} - \omega_\mu{}^b{}_c \sigma^{Ac}
    + \omega_\mu \epsilon^A{}_C \sigma^{Cb}
    + \sigma^b{}_c \omega_\mu{}^{Ac}
    - \sigma \epsilon^A{}_C \omega_\mu{}^{Cb}\,,
    \\
    \delta \pi^A_\mu
    &= \pd_\mu \kappa^A
    + \omega_\mu \epsilon^A{}_B \kappa^B
    - \sigma \epsilon^A{}_B \pi^B_\mu
    - \omega_\mu{}^A{}_b \kappa^b + \sigma^A{}_b \pi^b_\mu
    \\\nonumber
    &{}\qquad
    + \epsilon^A{}_B \omega_\mu{}^B{}_c \zeta^c
    - \epsilon^A{}_B \sigma^B{}_c e^c_\mu\,,
    \\
    \delta \pi^a_\mu
    &= \pd_\mu \kappa^a
    - \omega_\mu{}^a{}_b \kappa^b + \sigma^a{}_b \pi^b_\mu
    + \epsilon_{BC} \tau^B_\mu \sigma^{Ca}
    - \epsilon_{BC} \zeta^B \omega_\mu{}^{Ca}\,.
  \end{align}
\end{subequations}
The transformation parameters $\sigma$, $\sigma^{Ab}$ and $\sigma^{ab}=-\sigma^{ba}$ correspond to local Lorentz boosts, string Galilei boosts and transverse rotations, which we want to retain in the resulting TSNC geometry.
In contrast, we want to exchange the local translations $\zeta^A$ and $\zeta^a$ for diffeomorphisms $\xi^\mu$.
Likewise, we want to exchange the transformations $\kappa^A$ and $\kappa^a$ for the one-form gauge transformations
\begin{equation}
  \label{eq:app-m-mu-nu-gauge-tr-no-delta}
  m_{\mu\nu} \to m_{\mu\nu} + 2 \pd_{[\mu} \lambda_{\nu]}.
\end{equation}
To achieve this, we will now construct a modified transformation (denoted by $\bar\delta$) from the $\delta$-transformations in Equation~\eqref{eq:tsnc-adjoint-transformations}.
We will also need the expressions
\begin{subequations}
  \label{eq:app-tsnc-curvatures}
  \begin{align}
    \label{eq:app-tsnc-curvatures-H-A}
    R(H)^A
    &= d\tau^A + \epsilon^A{}_B \omega \wedge \tau^B\,,
    \\
    R(P)^a
    &= de^a - \omega^a{}_b \wedge e^b + \omega^{Ba} \wedge \tau_B\,,
    \\
    R(J)
    &= d\omega\,,
    \\
    R(J)^{ab}
    &= d\omega^{ab} - \omega^a{}_c \wedge \omega^{cb}\,,
    \\
    R(G)^{Ab}
    &= d\omega^{Ab} + \epsilon^A{}_C \omega \wedge \omega^{Cb}
    - \omega^b{}_c \wedge \omega^{Ac}\,,
    \\
    R(N)^A
    &= d \pi^A + \epsilon^A{}_B \omega \wedge \pi^B
    - \omega^A{}_b \wedge \pi^b
    + \epsilon^A{}_B \omega^B{}_c \wedge e^c\,,
    \\
    R(Q)^a
    &= d\pi^a - \omega^a{}_b \wedge \pi^b
    + \epsilon_{BC} \omega^{Ba} \wedge \tau^C\,,
  \end{align}
\end{subequations}
which correspond to the components $R(T)^A T_A$, with $T_A$ the generators of the Lie algebra, of the curvature $F = d\mA + \mA \wedge \mA$.

\paragraph{Vielbein and spin connection transformations}
We first define the $\bar\delta$ transformations on the vielbeine and spin connections, where we want to reproduce the Lie derivatives using a straightforward modification of the prescription in~\cite{Hartong:2015zia}.
The diffeomorphism generator $\xi^\mu$ is related to the local translations by
\begin{equation}
  \xi^\mu
  = v^\mu_A \zeta^A + \theta^\mu_a \zeta^a,
  \qquad
  \zeta^A = \tau^A_\mu \xi^\mu,
  \quad
  \zeta^a = e^a_\mu \xi^\mu,
\end{equation}
where $v_A^\mu$ and $\theta_a^\mu$ are the longitudinal and transverse inverse vielbeine, respectively.
These vielbeine satisfy the completeness and orthonormality conditions
\begin{equation}
  \label{eq:app-inverse-vielbeine-def}
  \delta^\mu_\nu
  = v^\mu_A \tau_\nu^A + \theta^\mu_a e_\nu^a\,,
  \qquad
  v^\mu_A e_\mu^b = 0\,,
  \quad
  \theta^\mu_a \tau_\mu^B = 0\,,
  \quad
  v^\mu_A \tau^B_\mu = \delta_A^B\,,
  \quad
  \theta^\mu_a e_\mu^b = \delta_a^b\,.
\end{equation}
For the longitudinal vielbeine, we define the transformation
\begin{align}
  \bar\delta\tau^A_\mu
  &= \LL_\xi \tau^A_\mu - \lambda \epsilon^A{}_B \tau^B\,.
\end{align}
This transformation contains diffeomorphisms and local longitudinal Lorentz boosts.
To show that this transformation can be constructed from the ingredients we have available from the gauging of the FSG algebra, we write it in terms of the adjoint transformation~\eqref{eq:tsnc-adjoint-transformations-tau-A} and a term involving the curvature~\eqref{eq:app-tsnc-curvatures-H-A},
\begin{align}
  \bar\delta\tau^A_\mu
  &= \xi^\nu\pd_\nu \tau^A_\mu + \tau^A_\nu \pd_\mu \left(
    v^\nu_B \zeta^B + \theta^\nu_b \zeta^b
  \right)
  - \lambda \epsilon^A{}_B \tau^B_\mu
  \\
  &= \pd_\mu \zeta^A - \lambda \epsilon^A{}_B \tau^B_\mu
  - \xi^\nu \left(
    \pd_\mu \tau^A_\nu - \pd_\nu \tau^A_\mu
  \right)
  \\
  &= \pd_\mu \zeta^A - \left(\lambda + \xi^\nu \omega_\nu\right)\epsilon^A{}_B\tau^B_\mu
  + \epsilon^A{}_B \omega_\mu \zeta^B
  - \xi^\nu R(H)_{\mu\nu}
  \\
  &= \delta \tau^A_\mu - \xi^\nu R(H)_{\mu\nu}\,.
\end{align}
For this, we identified the corresponding local Lorentz parameters as $\lambda + \xi^\rho\omega_\rho = \sigma$.
The $\bar\delta$~transformations for the spatial vielbeine $e^a_\mu$ and the spin connections are then defined as
\begin{align}
  \bar\delta e^a_\mu
  &= \LL_\xi e^a_\mu + \lambda^a{}_b e^b_\mu
  - \lambda_B{}^a \tau^B_\mu
  &&= \delta e^a_\mu - \xi^\nu R(P)^a_{\mu\nu}\,,
  \\
  \bar\delta \omega_\mu
  &= \LL_\xi \omega_\mu + \pd_\mu \lambda
  &&= \delta \omega_\mu - \xi^\nu R(J)_{\mu\nu}\,,
  \\
  \bar\delta \omega_\mu{}^{ab}
  &= \LL_\xi \omega_\mu{}^{ab} + \pd_\mu \lambda^{ab}
  - \omega_\mu{}^a{}_c \lambda^{cb}
  - \omega_\mu{}^b{}_c \lambda^{ac}
  &&= \delta \omega_\mu{}^{ab} - \xi^\nu R(J)_{\mu\nu}^{ab}\,,
  \\
  \bar\delta \omega_\mu{}^{Ab}
  &= \LL_\xi \omega_\mu{}^{Ab}
  + \pd_\mu \lambda^{Ab} + \lambda^b{}_c \omega_\mu{}^{Ac}
  - \epsilon^A{}_C \lambda \omega_\mu{}^{Cb}
  &&= \delta \omega_\mu{}^{Ab}
  - \xi^\nu R(G)^{Ab}_{\mu\nu}\,.
\end{align}
On the right hand side, a similar computation as for $\bar\delta\tau^A_\mu$ allows us to rewrite these $\bar\delta$~transformations in terms of the adjoint $\delta$ transformations in Equation~\eqref{eq:tsnc-adjoint-transformations}, supplemented by curvature terms.
The transformation parameters are related by
\begin{equation}
  \label{eq:gauging-fsnc-sigma-lambda-identification}
  \sigma = \lambda + \xi^\mu \omega_\mu\,,
  \quad
  \sigma^{ab} = \lambda^{ab} + \xi^\mu \omega_\mu{}^{ab}\,,
  \quad
  \sigma^{Ab} = \lambda^{Ab} + \xi^\mu \omega_\mu{}^{Ab}\,.
\end{equation}
Building on the adjoint transformations, and using appropriate curvature terms, we can construct a set of $\bar\delta$ transformations of the vielbeine and spin connections in which local translations are replaced by diffeomorphisms.
This allows us to define a fully diffeomorphism-covariant notion of torsional string Newton--Cartan (TSNC) geometry using the ingredients available from the gauging of the FSG algebra.

\paragraph{Kalb--Ramond-type fields}
However, we are still missing one key ingredient.
We want to reproduce the appropriate symmetry transformations for the two-form field~\eqref{eq:app-mdef-repeat} that is constructed out of the vielbeine and the gauge fields $\pi^A_\mu$ and $\pi^a_\mu$.
Diffeomorphisms can be constructed from local translations in a similar manner as we did for the vielbeine and spin connections above.

In addition, we want to reproduce the one-form gauge transformations~\eqref{eq:app-m-mu-nu-gauge-tr-no-delta}.
We now show that these can be constructed from the transformations in~\eqref{eq:tsnc-adjoint-transformations} parametrized by $\kappa^A$ and $\kappa^a$, supplemented with appropriate curvature terms.
Initially, these parameters act on $m_{\mu\nu}$ as follows,
\begin{equation}
  \delta_\kappa m_{\mu\nu}
  = 2 \pd_{[\mu} \lambda'_{\nu]}
  - \kappa_A R(H)^A_{\nu\mu}
  - \kappa_a R(P)^a_{\nu\mu}\,.
\end{equation}
Here, we have identified $2\lambda'_\mu = - \eta_{AB} \tau^A_\mu \kappa^B - \delta_{ab} e^a_\mu \kappa^b$.

To recover the desired one-form transformation~\eqref{eq:app-m-mu-nu-gauge-tr-no-delta}, one could demand that the torsion $R(H)^A_{\mu\nu}$ and $R(P)^a_{\mu\nu}$ vanishes.
However, this is not necessary.
If we can subtract these curvature terms by appropriately defining the $\bar\delta$ transformations of $\pi^A_\mu$ and $\pi^a_\mu$, we can get the correct one-form transformation without imposing any constraint on the torsion.
Simultaneously, we want to obtain the appropriate transformations under diffeomorphisms.
To do this, we define the transformations
\begin{subequations}
  \label{eq:gauging-fsnc-deltabar-pi-transformations}
  \begin{align}
    \bar\delta \pi^A_\mu
    &= \LL_\xi \pi^A_\mu + \pd_\mu k^A
    + \epsilon^A{}_B \omega_\mu k^B
    - \omega_\mu{}^A{}_b k^b
    \\\nonumber
    &{}\qquad
    + \lambda^A{}_b \pi^b_\mu
    - \epsilon^A{}_B \lambda \pi_\mu^B
    - \epsilon^A{}_B \lambda^B{}_c e^c_\mu
    + \frac{1}{2} k_B v^{A\nu} R(H)^B_{\mu\nu}
    + \frac{1}{2} k_b v^{A\nu} R(P)^b_{\mu\nu}\,,
    \\
    \bar\delta \pi^a_\mu
    &= \LL_\xi \pi^a_\mu + \pd_\mu k^a
    - \omega_\mu{}^a{}_b k^b + \lambda^a{}_b \pi^b_\mu
    + \epsilon_{BC} \tau^B_\mu \lambda^{Ca}
    \\\nonumber
    &{}\qquad
    + \frac{1}{2} k_b \theta^{a\nu} R(P)^b_{\mu\nu}
    + \frac{1}{2} k_B \theta^{a\nu} R(H)^B_{\mu\nu}\,.
  \end{align}
\end{subequations}
As before, these can be written in terms of the adjoint transformations and curvatures,
\begin{subequations}
  \begin{align}
    \bar\delta \pi^A_\mu
    &= \delta \pi^A_\mu - \xi^\nu R(N)^A_{\mu\nu}
    + \frac{1}{2} k_B v^{A\nu} R(H)^B_{\mu\nu}
    + \frac{1}{2} k_b v^{A\nu} R(P)^b_{\mu\nu}\,,
    \\
    \bar\delta \pi^a_\mu
    &= \delta \pi^a_\mu - \xi^\nu R(Q)^a_{\mu\nu}
    + \frac{1}{2} k_b \theta^{a\nu} R(P)^b_{\mu\nu}
    + \frac{1}{2} k_B \theta^{a\nu} R(H)^B_{\mu\nu}\,,
  \end{align}
\end{subequations}
where $v^\mu_A$ and $\theta^\mu_a$ are the inverse vielbeine defined by Equation~\eqref{eq:app-inverse-vielbeine-def}.
Here, in addition to~\eqref{eq:gauging-fsnc-sigma-lambda-identification} relating the $\sigma$ and $\lambda$ parameters, we identify
\begin{equation}
  \kappa^A = k^A + \xi^\mu \pi_\mu^A\,,
  \quad
  \kappa^a = k^a + \xi^\mu \pi_\mu^a\,.
\end{equation}
The two-form field $m_{\mu\nu}$ from Equation~\eqref{eq:app-mdef-repeat} is invariant under local boosts and rotations.
The correct transformation of $m_{\mu\nu}$ under diffeomorphisms follows from that of $\pi^A_\mu$, $\pi^a_\mu$ and the vielbeine $\tau^A_\mu$ and $e^a_\mu$.
Finally, we see that the $\bar\delta$ transformations in~\eqref{eq:gauging-fsnc-deltabar-pi-transformations} result in the desired one-form gauge transformation
\begin{equation}
  \label{eq:fsnc-b-field-transformation-deltabar}
  \bar\delta m_{\mu\nu} = 2 \pd_{[\mu} \lambda_{\nu]}\,,
\end{equation}
where the one-form transformation parameter $\lambda_\mu$ is given by
\begin{equation}
  \label{eq:gauging-fsnc-lambda-k-relation}
  \lambda_\mu
  = - \frac{1}{2} \left( \eta_{AB} \tau^A_\mu k^B + \delta_{ab} e^a_\mu k^b \right)\,.
\end{equation}
With this, we have obtained both the diffeomorphisms and one-form gauge transformations of TSNC geometry from the gauging of the FSG algebra, without any torsion constraints.

\bibliographystyle{JHEP}
\bibliography{biblio}

\end{document}